\documentclass[twocolumn,times]{aastex6}

\usepackage{amsfonts}
\usepackage{mathrsfs}
\usepackage{rotating}

\newcommand{\fagn}{$F_{\rm AGN}$}
\newcommand{\fcs}{$F_{\rm 5100}$}
\newcommand{\feii}{Fe {\sc ii}}
\newcommand{\ffe}{$F_{\rm Fe}$}
\newcommand{\fhb}{$F_{\rm H\beta}$}
\newcommand{\fhbf}{$F_{\rm H\beta,fit}$}
\newcommand{\fhbs}{$F_{\rm H\beta,intg}$}
\newcommand{\fhei}{$F_{\rm He~\textsc{i}}$}
\newcommand{\fheis}{$F_{\rm He~\textsc{i},intg}$}
\newcommand{\fheii}{$F_{\rm He~\textsc{ii}}$}
\newcommand{\fvar}{$F_{\rm var}$}
\newcommand{\hb}{H$\beta$}
\newcommand{\hei}{He {\sc i}}
\newcommand{\heii}{He {\sc ii}}
\newcommand{\mbh}{$M_\bullet$}
\newcommand{\oiii}{[O~{\sc iii}]}
\newcommand{\rmax}{$r_{\rm max}$}

\def\tauhb{\tau_{\rm H\beta}}
\def\kms{km\,${\rm s^{-1}}$}
\def\mcn{\multicolumn}
\def\mathdotM{\dot{\mathscr{M}}}
\def\ergs{\rm erg\,s^{-1}}
\def\ergscm{\rm erg\,s^{-1}\,cm^{-2}}
\def\ergscma{\rm erg\,s^{-1}\,cm^{-2}\,\AA^{-1}}
\def\fblr{f_{\rm BLR}}
\def\mbh{M_{\bullet}}
\def\sunm{M_{\odot}}

\slugcomment{Accepted for publication in \textit{The Astrophysical Journal}}

\shorttitle{Broad-line Region of PG 2130+099}
\shortauthors{Hu et al.}

\begin{document}

\title{Broad-line Region of the Quasar PG 2130+099 from a Two-Year
Reverberation Mapping Campaign with High Cadence}

\author
{Chen Hu\altaffilmark{1}, 
Yan-Rong Li\altaffilmark{1},
Pu Du\altaffilmark{1},
Zhi-Xiang Zhang\altaffilmark{1,2},
Sha-Sha Li\altaffilmark{1,2},
Ying-Ke Huang\altaffilmark{1,2}, 
Kai-Xing Lu\altaffilmark{3},
Jin-Ming Bai\altaffilmark{3},
Luis C. Ho\altaffilmark{4,5},
Wei-Hao Bian\altaffilmark{6},
Michael S. Brotherton\altaffilmark{7},
Ye-Fei Yuan\altaffilmark{8},
Jes\'us Aceituno\altaffilmark{9,10}, \\
Hartmut Winkler\altaffilmark{11} and
Jian-Min Wang\altaffilmark{1,2,12}\\
(SEAMBH collaboration)}

\altaffiltext{1}
{Key Laboratory for Particle Astrophysics, Institute of High Energy Physics, 
The Chinese Academy of Sciences, 19B Yuquan Road, Beijing 100049, China; 
huc@ihep.ac.cn, wangjm@ihep.ac.cn}

\altaffiltext{2}
{School of Astronomy and Space Science, University of Chinese Academy of
Sciences, 19A Yuquan Road, Beijing 100049, China}

\altaffiltext{3}
{Yunnan Observatories, The Chinese Academy of Sciences, Kunming 650011, China}

\altaffiltext{4}
{Kavli Institute for Astronomy and Astrophysics, Peking University, Beijing
100871, China}

\altaffiltext{5}
{Department of Astronomy, School of Physics, Peking University, Beijing
100871, China}

\altaffiltext{6}
{Physics Department, Nanjing Normal University, Nanjing 210097, China}

\altaffiltext{7}
{Department of Physics and Astronomy, University of Wyoming, Laramie, WY
82071, USA}

\altaffiltext{8}
{Department of Astronomy, University of Science and Technology of China, Hefei
230026, China}

\altaffiltext{9}
{Centro Astronomico Hispano Alem\'an, Sierra de los filabres sn, 04550 gergal.
Almer\'ia, Spain}

\altaffiltext{10}
{Instituto de Astrof\'isica de Andaluc\'ia, Glorieta de la astronom\'ia sn,
18008 Granada, Spain}

\altaffiltext{11}
{Department of Physics, University of Johannesburg, PO Box 524, 2006 Auckland
Park, South Africa}

\altaffiltext{12}
{National Astronomical Observatories of China, The Chinese Academy of
Sciences, 20A Datun Road, Beijing 100020, China}

\begin{abstract}
As one of the most interesting Seyfert 1 galaxies, PG 2130+099 has been the
target of several reverberation mapping (RM) campaigns over the years.
However, its measured broad H$\beta$ line responses have been inconsistent,
with time lags of $\sim$200 days, $\sim$25 days, and $\sim$10 days being
reported for different epochs while its optical luminosity changed no more
than 40\%. To investigate this issue, we conducted a new RM-campaign with
homogenous and high cadence (about $\sim$3 days) for two years
during 2017--2019 to measure the kinematics and structure of the ionized
gas. We successfully detected time lags of broad \hb, \heii, \hei, and
\feii\ lines with respect to the varying 5100\AA\ continuum, 
revealing a stratified structure that is likely virialized with Keplerian
kinematics in the first year of observations, but an inflow kinematics
of the broad-line region from the second year. With a central black hole mass
of $0.97_{-0.18}^{+0.15}\times 10^7~\sunm$, PG 2130+099 has an
accretion rate of $10^{2.1\pm0.5}L_{\rm Edd}c^{-2}$, where $L_{\rm Edd}$ is
the Eddington luminosity and $c$ is speed of light, implying that it is a
super-Eddington accretor and likely possesses a slim, rather than thin,
accretion disk. The fast changes of the ionization structures of the
three broad lines remain puzzling.
\end{abstract}


\section{Introduction}

Reverberations, the responses of broad emission lines to a changing continuum
\citep{Bachall1972,blandford82}, convey invaluable information about the
kinematics and structure of the broad-line regions (BLRs) 
of Seyfert 1 galaxies \citep[for a review]{peterson14}.  As an outlier of the
well-known $R-L$ relation discussed by \citet{Bentz2013}, the Seyfert 1 galaxy
PG 2130+099 ($m_V=14.3$, $z=0.063$) has drawn much attention in recent years.
This intriguing object is a candidate super-Eddington accreting massive black
hole (SEAMBH), as it shares the weak \oiii\, lines, strong \feii\ emission,
and Lorentzian broad \hb\ profile with others of the class \citep{du15}. 
However, the reported \hb\ time lags are mostly inconsistent with the 
characteristically short lags of SEAMBHs.

\citet{kaspi00} obtained a broad \hb\ line time lag of
$\tauhb=188_{-27}^{+136}$ days from nearly eight-years monitoring between 
1990 and 1998, but their campaign had very low cadence (the median sampling is 
$\gtrsim 20$ days). In a 2007 campaign, \citet{grier08} measured a lag of
$\tauhb=22.9_{-4.3}^{+4.4}$ days,  which is roughly an order of magnitude
shorter. They claimed that the data of Kaspi et al. (2000) is generally too
undersampled to measure a short lag, and that the $\tauhb=188$ days could be
caused by the secular long-term variability of the \hb\ line. However,
the time span of the \citet{grier08} data is also too short (only $\sim$100
days) to measure a long lag, and the cadence is not uniform, while the number
of epochs is rather small as well (only 21). 

\citet{grier12} monitored PG 2130+099 once again in 2010, with much better
sampling of $\sim$1 days in median and a longer campaign duration of $\sim$120
days, but the \hb\ lag again differed from previous investigations.  They
obtained $\tauhb=12.8_{-0.9}^{+1.2}$ days from the likelihood method, and
$\tauhb=9.6\pm1.2$ days from the CCF methods with $r_{\rm max}$ only
$\sim$0.35. \citet{Grier2013b} obtained a velocity-resolved delay for the
2010 campaign, but only four velocity bins were good enough for \hb\ lags (see
their Figure 7), and there are two peaks ($\sim$10 and $\sim$30 days, and
even a third at $\sim$50 days at the very edge) in the transfer function
(see their Figure 15). Their 2-dimensional transfer function shows the
dominant response with $\sim$30 days from the maximum-entropy method (see
their Figure 12). \citet{Grier2013b} note that their campaign missed some
potentially important epochs, and that this affected the accuracy of their
results. The \hb\ lags are complex although \citet{Bentz2013} prefers
$\tauhb\approx 30$ days. \citet{Grier2017} calculated a dynamical model for
the \hb-emitting region that is a thick Keplerian disk, but their model does
not match the late 1/3 of the entire light curve (see their right panel of
Figure 2), and seems to ignore the inflow component clearly visible 
in the results of \citet{Grier2013b}. 

Once again, this enigmatic object requires additional observations to
determine its behavior, BLR structure and \hb\ time lag. In this
paper, we report results of a spectral monitoring campaign of PG 2130+099
mainly based on observations with the the Centro Astron\'omico
Hispano-Alem\'an (CAHA) 2.2m telescope at the Calar Alto Observatory in Spain.
We successfully detected reverberations of several broad emission lines and
thus have robust measurements of the BLR. Implications for our understanding
of the BLR are discussed concerning the presence of multiple sub-structures.
We use a cosmology with 
$H_0=70\,{\rm km\,s^{-1}\,Mpc^{-1}}$ and $\Omega_{\rm M}=0.3$.

\section{Observations and data reduction}

Since May 2017, the SEAMBH project \citep{du14} has expanded to make use of
the CAHA 2.2m telescope to perform  measurements focusing on PG quasars with
high accretion rates (hereafter the CAHA-SEAMBH project). Given the
uncertainties about its time lag and mass, PG 2130+099 is one of highest
priorities of the CAHA-SEAMBH project. The monitoring 
was carried out at the telescope with the Calar Alto Faint Object Spectrograph
(CAFOS) for 53 nights between 2017 June 22 (hereafter the year of 2017
observations) and 2018 January 4, and 58 nights between 2018 May 31
and 2019 January 16 (hereafter the year of 2018 observations). The strategy
of observations is similar to that of the SEAMBH campaign performed at the
Lijiang 2.4m telescope of the Yunnan Observatories, Chinese Academy of
Sciences \citep{du14}. For each night, broadband images
of the object were obtained, then a long slit with a projected width of
3$\farcs$0 was carefully oriented and the spectra of the object and a nearby
non-varying comparison star were taken simultaneously. The typical
displacement between the object and comparison star perpendicular to the slit
direction was less than 0.5 pixel (0$\farcs$265), meaning that both are
similarly affected by seeing losses, and thus enabling high-accurate flux
calibration (the typical seeing was less than 2 arcsec). The position angle of
the comparison star is $248^{\circ}$, and its angular separation from
PG2130+099 is $86\arcsec$. The Sloan Digital Sky Survey
(SDSS; \citealt{york00}) gives the magnitude of the comparison star in
$\it g$ band as 15.1. Our spectra show the comparison to be a G-type star.
The typical S/N of the comparison star is $\sim$50 per pixel around 5200\AA,
corresponding to the \hb\ wavelength range of PG 2130+099 in the observed
frame, which is sufficient for our requirement of flux calibration. We also
restricted the airmass in most observations to be less than 1.2
to minimize the effect of atmospheric differential refraction.

\subsection{Photometry}

The photometric images were obtained by CAFOS in direct imaging mode with a
Johnson $V$ filter. Typically, three exposures of 20 s were taken each
individual night. The images were reduced following standard IRAF procedures.
The flux of the object and the comparison star were measured through a circular
aperture with radius of $2\farcm1$, and differential magnitudes were obtained
relative to 17 other stars within the field of 8$\farcm$8$\times$8$\farcm$8.

\begin{figure*}
  \centering
  \includegraphics[width=0.75\textwidth]{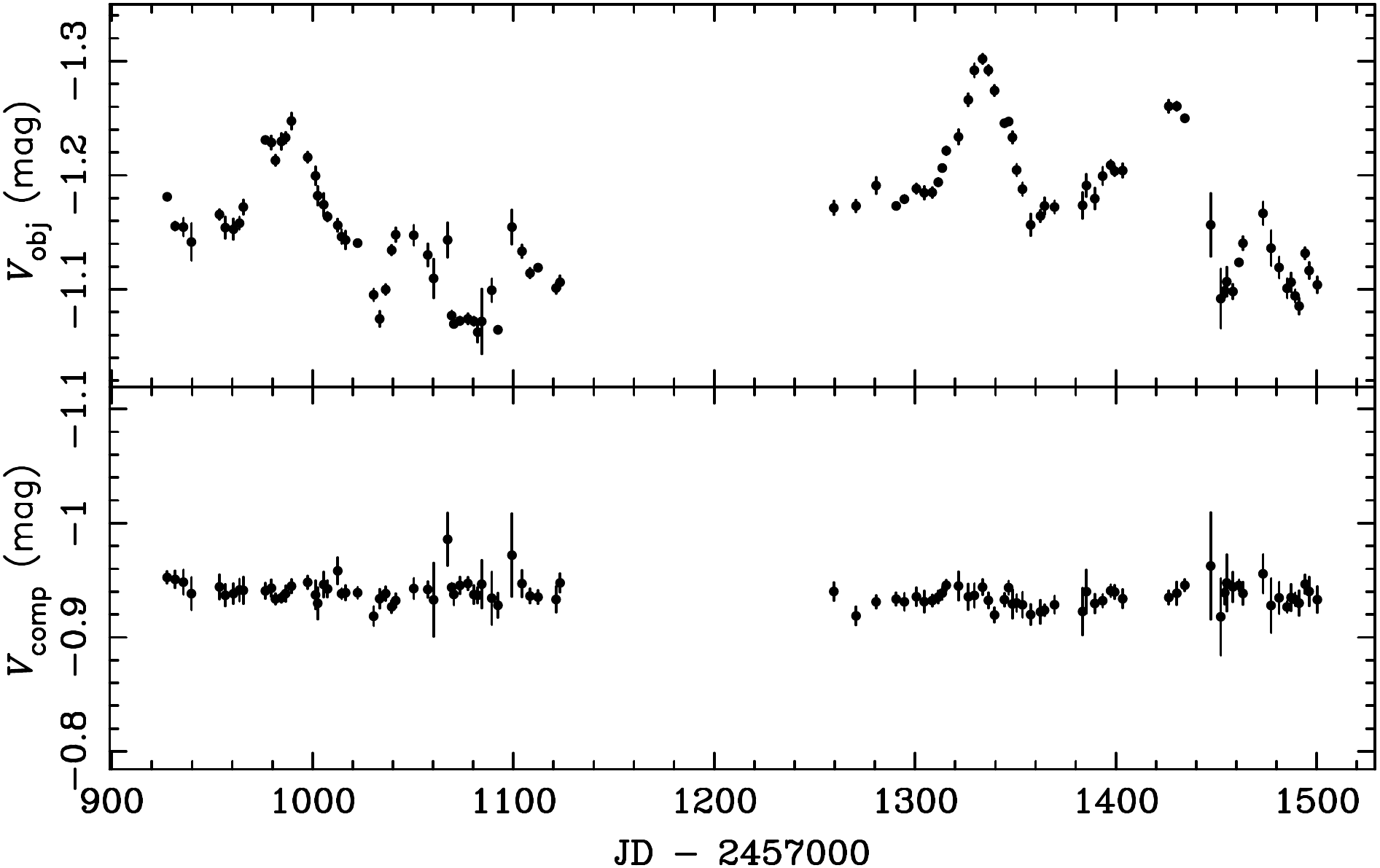}
  \caption{\footnotesize Light curves in the $V$ band for PG 2130+099
  (top) and the comparison star (bottom). The scatter in the magnitudes of the
  comparison star is $\sim$0.01 mag.
  }
  \label{fig-lcphot}
\end{figure*}

Figure \ref{fig-lcphot} shows the light curves of the differential magnitudes
for the object (top) and the comparison star (bottom), respectively. The
accurate of the photometry is $\sim$0.01 mag, estimated from the scatter in the
magnitudes of the comparison star. It also shows that this comparison star is
stable enough to be used for the flux calibration of the spectroscopy.

\subsection{Spectroscopy}

The spectra of PG 2130+099 were taken using CAFOS with Grism G-200, which has
a relatively high efficiency, without an order-blocking filter.  The resulting
spectra cover the observed-frame wavelength range of 4000--8500 \AA, with a
dispersion of 4.47 \AA\ pixel$^{-1}$. Note that the second-order contamination
occurs for wavelengths longer than $\sim$7000 \AA, and has no effect on our
the measurements.  A set of calibration frames was taken for each night,
including bias frames, dome flats, and wavelength-calibration lamps of
HgCd/He/Rb. For the object, two exposures of 600 s were taken during each
individual night, and the typical signal-to-noise ratio (S/N) of the spectrum
for a single exposure is $\sim$80 per pixel at rest-frame 5100 \AA. We took
one or two spectrophotometric standards in the
dusk and/or dawn of each night, if the weather and time allowed.

The spectra were reduced with IRAF following standard procedures:
bias-removal, flat-fielding, wavelength calibration, and extraction to one
dimension.  Extraction apertures were uniform and  large (10$\farcs$6) to
minimize light loss, especially on nights with poor seeing. The flux
calibration of the object used the sensitivity function determined from the
comparison star, as described below.  Firstly, for several nights with good
weather conditions, the normal IRAF procedure of flux calibration was
performed for the comparison star using spectrophotometric standards taken on
the same night. 
Secondly, the calibrated spectra for those nights were combined to generate a
fiducial spectrum of the comparison star. Then, for each exposure, a Legendre
polynomial was fitted by comparing the extracted spectrum of the comparison
star in counts to the fiducial spectrum. This Legendre polynomial serves as
the sensitivity function and incorporates all corrections including the
atmosphere, slit loss, and instrumental sensitivity. Finally, the spectrum of
the object for each exposure was flux-calibrated using the corresponding
sensitivity function, and the two calibrated spectra for the same night were
combined to obtain the individual-night spectrum for the following
measurements and analysis.

\begin{figure}
  \centering
  \includegraphics[width=0.45\textwidth]{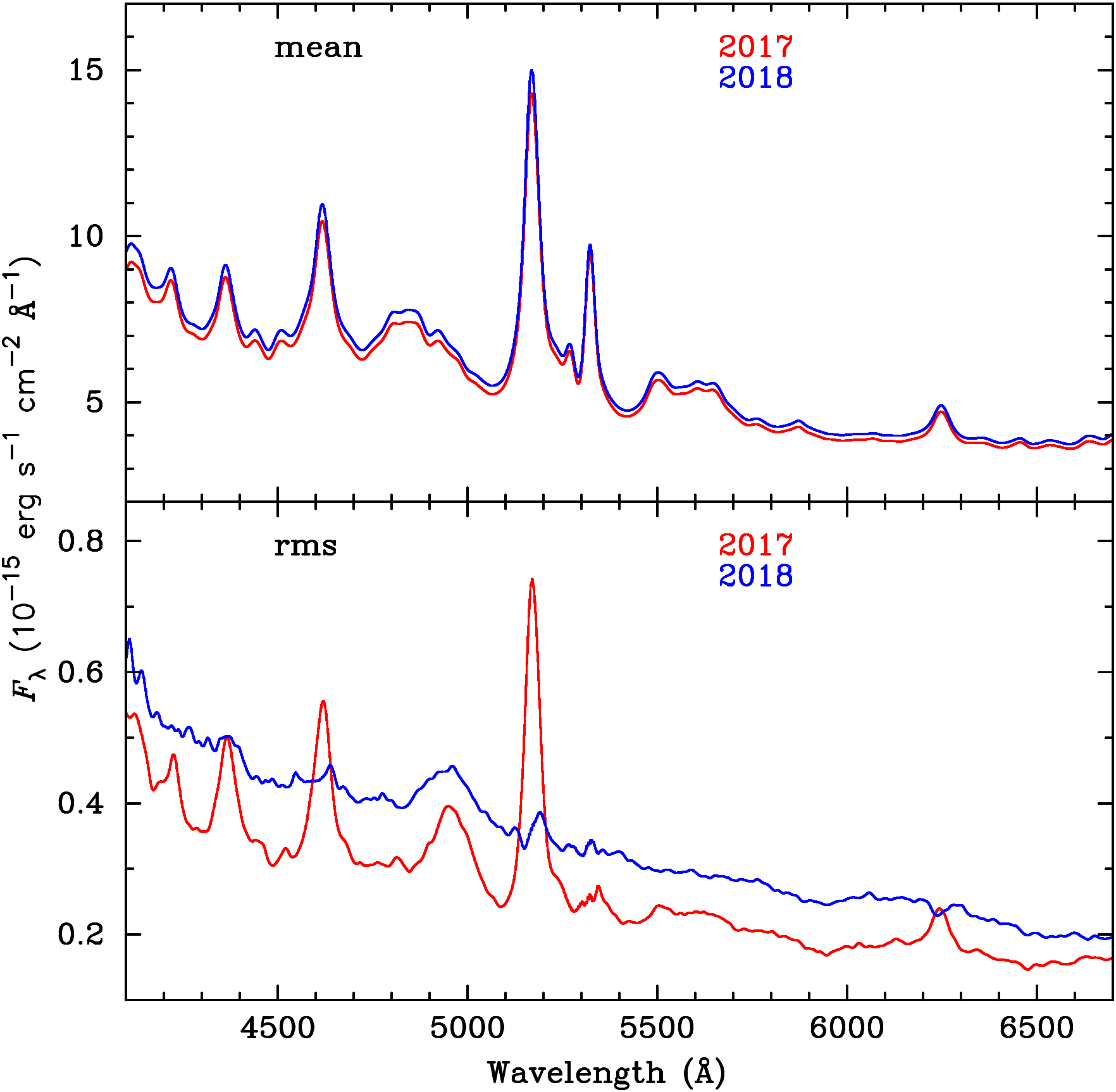}
  \caption{\footnotesize The mean (top) and rms (bottom) spectra for
  years 2017 (red) and 2018 (blue) separately. The mean spectra are almost the
  same, while the rms spectra are surprisingly different in the Balmer lines.
  }
  \label{fig-rms}
\end{figure}

The accuracy of the flux calibration using our comparison stars has been
proven to be better than $\sim$3\% \citep{du18}. In this specific case, it is
better than 3\% as estimated from the scatter of the \oiii\ light curve (see
Section \ref{sec-sfs} for details). Figure \ref{fig-rms} shows the
mean (top) and RMS (bottom) spectra for the years 2017 (red) and 2018 (blue)
separately. The \oiii\ line almost vanishes in the rms spectrum, indicating
that we have achieved good flux calibration. A prominent feature of the rms
spectrum, compared to the mean spectrum, is the strong, broad \heii\ emission
line, indicating the strong variation in this line. The mean spectra
for the two years are almost identical, and the variations of the \hb\
profiles are also quite small compared with those in \cite{kaspi00} and
\cite{grier12}. However, the rms spectrum of 2018 is dramatically different in
its almost vanishing Balmer lines.

\subsection{Other telescopes}

In 2018, the object was also monitored by two other telescopes: Lijiang 2.4m
telescope and Sutherland 1.9m telescope. Spectra were taken on three nights
at the Lijiang 2.4m telescope at the Yunnan Observatory of the Chinese Academy
of Sciences, using the Yunnan Faint Object Spectrograph and Camera with grism
G14 and a long slit of $2\farcs5$. The slit was oriented to include the same
comparison star as at CAHA, which is used for flux calibration. The spectra
were extracted in an aperture of $2\farcs5\times8\farcs5$. The details of the
spectroscopy and data reduction are given in \citet{du14}. At the Sutherland
station of the South African Astronomical Observatory, spectroscopic
observations of PG 2130+099 were carried out with the 600 lines mm$^{-1}$
grating and a slit of $0\farcs9$, on 12 nights. The flux calibration was done
by scaling the flux of the narrow \oiii\ emission lines to be the same as the
mean value in the CAHA spectra. The set of the observations and data reduction
are as the same as that described in \citet{Winkler2017}.

The epochs of the Lijiang spectra are rather few (only 3 nights), while the
Sutherland spectra are calibrated by another method and the data quality is
relatively scattered especially for the last four data points where technical
problems caused a shift in the grating angle. So for the consistency of the
data set, we used the data from CAHA only for the analysis below, but show the
integrated light curves of the continuum and \hb\ from Lijiang (green points)
and Sutherland (blue points) in panels (a) and (c) of Figure \ref{fig-lc}.

\section{Analysis methods}

\subsection{Light curve measurements}

There are several methods to measure the light curves. The traditional method
adopted in most RM studies \citep[e.g.][]{peterson04,kaspi00,du18} employs
integration over a range of relevant wavelengths, which is simple and usually
robust for single, strong emission lines such as \hb. But for those highly
blended emission lines, e.g., \feii\ emission which form a pseudo-continuum,
a spectral-fitting scheme (SFS) has proven to be necessary to measure the
light curves \citep{bian10,barth13,hu15}. In some cases, especially for nearby
Seyferts, spectral fitting can improve the measurements for even the \hb\
emission line, by modeling and removing the contamination of strong host
galaxy contribution which varies from night to night due to seeing and guiding
variations \citep{hu15,hu16}. However, comparing with integration,
spectral fitting scheme requires higher quality of the data. For example, for
reliable decomposition of the host starlight, the wavelength coverage of the
spectrum should be wide, and the shape of the spectrum has to be especially
very well calibrated. These requirements can be difficult
to achieve with the \oiii-calibration 
mode of RM-campaigns since only a constant scaling factor is
applied for the whole wavelength range. Sometimes, integration and fitting are
combined to take the advantages of both methods: subtracting the continuum by
spectral fitting and then integrating the residual flux to measure the light
curves of emission lines \citep{barth13,barth15}.

For fitting the spectra of PG 2130+099 here, besides an AGN power law, a host
starlight template is needed to model the continuum shape accurately.
However, the host starlight contribution is rather weak in this luminous PG
object, and also the spectral resolution is low; our spectra do not show
prominent absorption features of host starlight. Thus, the host starlight
component can not be modeled well and its apparent variation due to seeing and
mis-centering cannot be removed by spectral fitting as precisely as for other
sources \citep{hu15,hu16}. In fact, the uncertainty in fitting the host
starlight could introduce extra systematic errors into the light curve
measurements. So in this work, we used both methods in this manner:
integration for the measurements of the AGN continuum, \hb, and \hei, and the
fitting method for the highly blended \feii\ and \heii\ features.

\subsubsection{Integration scheme}

Taking the following windows of H$\beta: 4810-4910$\AA\ and \hei:
$5950-6000$\AA\ in the rest frame, respectively, we integrate across them for
the \hb\ and \hei\ lines.
For each emission line, a straight line was defined by the two continuum
windows located on the sides of the emission, and then the flux above this
straight line was integrated in the emission-line window. The average flux in
the redward continuum window of \hb, 5085--5115 \AA, was calculated as the
integrated 5100 \AA\ flux (\fcs). Note that the individual-night spectra
were corrected for the Galactic extinction and de-redshifted before the
integration, to enable the comparison of the light curves given by
integration and fitting. We assumed an extinction law with $R_V$ = 3.1
\citep{cardelli89,odonnell94} and adopted a $V$-band extinction of 0.122 mag
obtained from the NASA/IPAC Extragalactic Database of \citet{schlafly11}.
Light curves are given in Figure \ref{fig-lc}, panels (a)-(c).

\subsubsection{Spectral-fitting scheme}
\label{sec-sfs}

\begin{figure}
  \centering
  \includegraphics[angle=-90,width=0.45\textwidth]{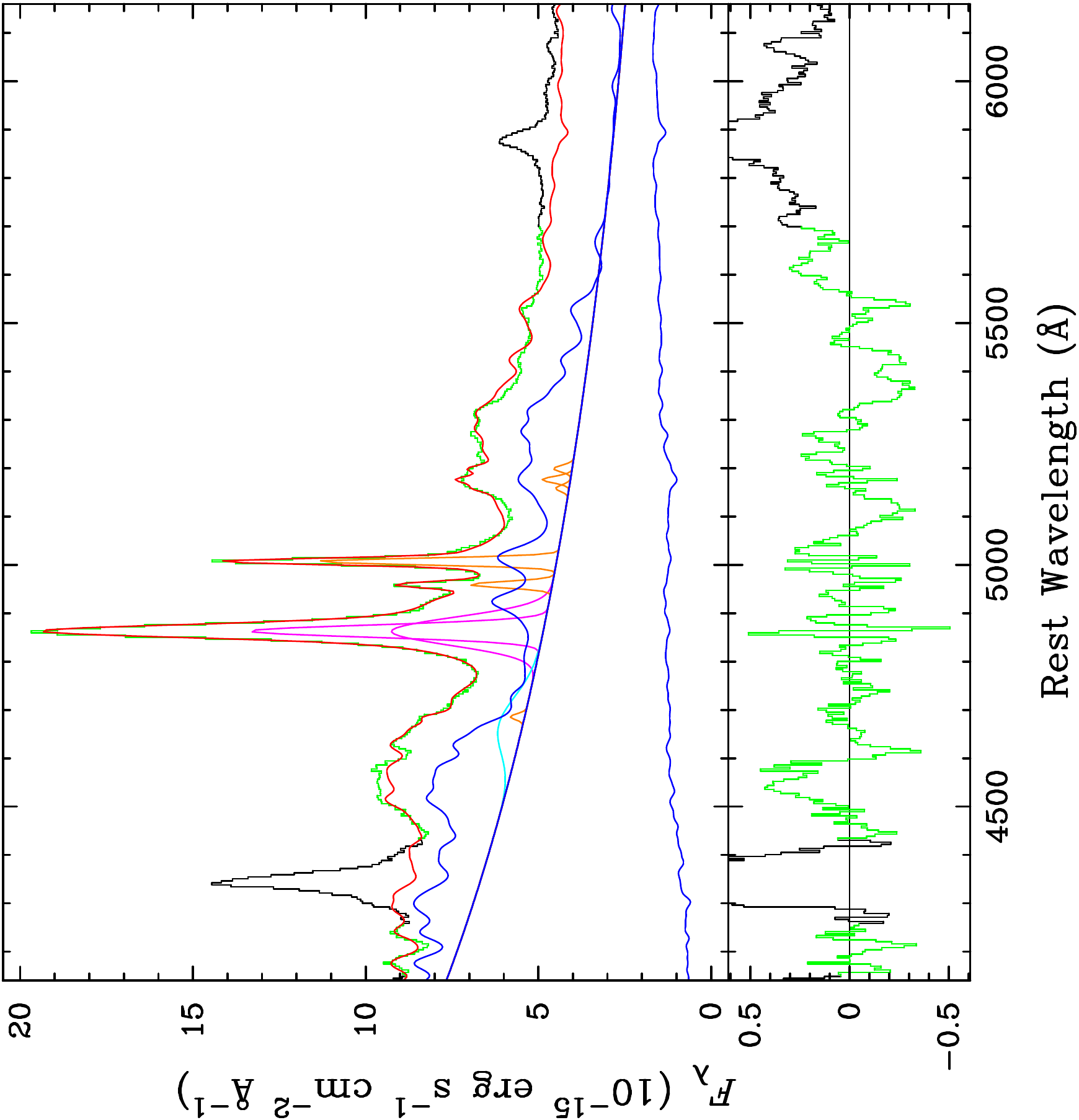}
  \caption{\footnotesize Spectral fitting scheme for  an individual-night
  spectrum. The top panel shows the calibrated spectrum (green for
  fitted pixels and black for excluded) and the best-fit model (red), composed
  by the AGN power-law continuum (blue), \feii\ emission (blue), host galaxy
  (blue), broad \hb\ (magenta), broad \heii\ (cyan), and narrow emission lines
  (orange). The bottom panel shows the residuals. Details can be found in the
  main text.
  }
  \label{fig-spec}
\end{figure}

The SFS follows \citet{hu15} with some changes described below to suit the
difficulty of decomposing the host starlight in our spectra. Figure
\ref{fig-spec} shows an example of the fit to an individual-night spectrum.
The following components are included in our fitting: (1) a single power law
for the AGN continuum, (2) \feii\ emission modeled by convolving the template
from \citep{bg92} with a Gaussian function, (3) the host galaxy starlight
modeled by the template with 11 Gyr age and metallicity $Z$ = 0.05 from
\citet{bc03}, (4) a double Gaussian for \hb, (5) a single Gaussian for \heii,
(6) a set of several single Gaussians with the same velocity width and shift
for narrow-emission lines including \oiii\ $\lambda\lambda$4959, 5007, \heii\
$\lambda$4686, and several coronal lines. The narrow \hb\ line is not
included in the fitting because of its weakness ($\sim$2\% of the total \hb\
flux estimated using \oiii\ line, see Sec. \ref{sec-lags} for details) and the
low spectral resolution prevent a realiable decomposition of it in
individual-night spectra. Thus, the narrow \hb\ line is ignored and the double
Gaussian of component (4) above are used for the measurements of broad \hb.
The \heii\ lines are weak
and blended with \feii\ emission in individual-night spectra and even the mean
spectrum, but prominent in the rms spectrum. Thus the velocity width and shift
of \heii\ are fixed to the values given by the best fit to the rms spectrum.
Also, the coronal lines are weak and blended with \feii\ emission, thus the flux 
ratios of the narrow-emission lines relative to \oiii\
$\lambda$5007 are fixed to the values given by the best fit to the mean
spectrum, as in \citet{hu15}. The only difference between the fitting here and
that in \citet{hu15} is the treatment of the host galaxy. The spectra of PG
2130+099 do not show prominent absorption features because of the weakness
of its host galaxy and also the low spectral resolution 
(with an instrumental broadening of $\sim$ 1000 \kms\ in FWHM). So the 
velocity width
and shift of the host galaxy are fixed to those of \oiii\ $\lambda$5007 in the
best fit to the mean spectrum. The flux of the host starlight is allowed to
vary, but the slope of the AGN continuum is fixed to the best-fit value of the
mean spectrum, as in \citet{hu15}. In total, there are 15 free parameters, the
other 17 are fixed.

The fitting is performed in the wavelength range 4150--5700 \AA, excluding a
narrow window around H$\gamma$. We try to extend the fit redward to include
the \hei\ $\lambda$5876 emission line. However, as shown as the residual in
Figure \ref{fig-spec}, this emission line seems to have a very broad wing, or
the continuum around it is not well modeled due to the uncertain host galaxy
component. Including this wavelength range in the fitting makes the resulting
continuum light curve more uncertain, that the systematic error (estimated by
the scatter in the fluxes of successive nights) becomes larger. Also, the light
curve of the fitted \hei\ resembles that from the integration method, but with
slightly larger scatter. So we
do not include \hei\ $\lambda$5876 in the fitting, and adopt its light curve
from direct integration.

\begin{sidewaysfigure*}
  \centering
  \vspace{-0.4\textwidth}
  \includegraphics[angle=-90,width=1.02\textwidth]{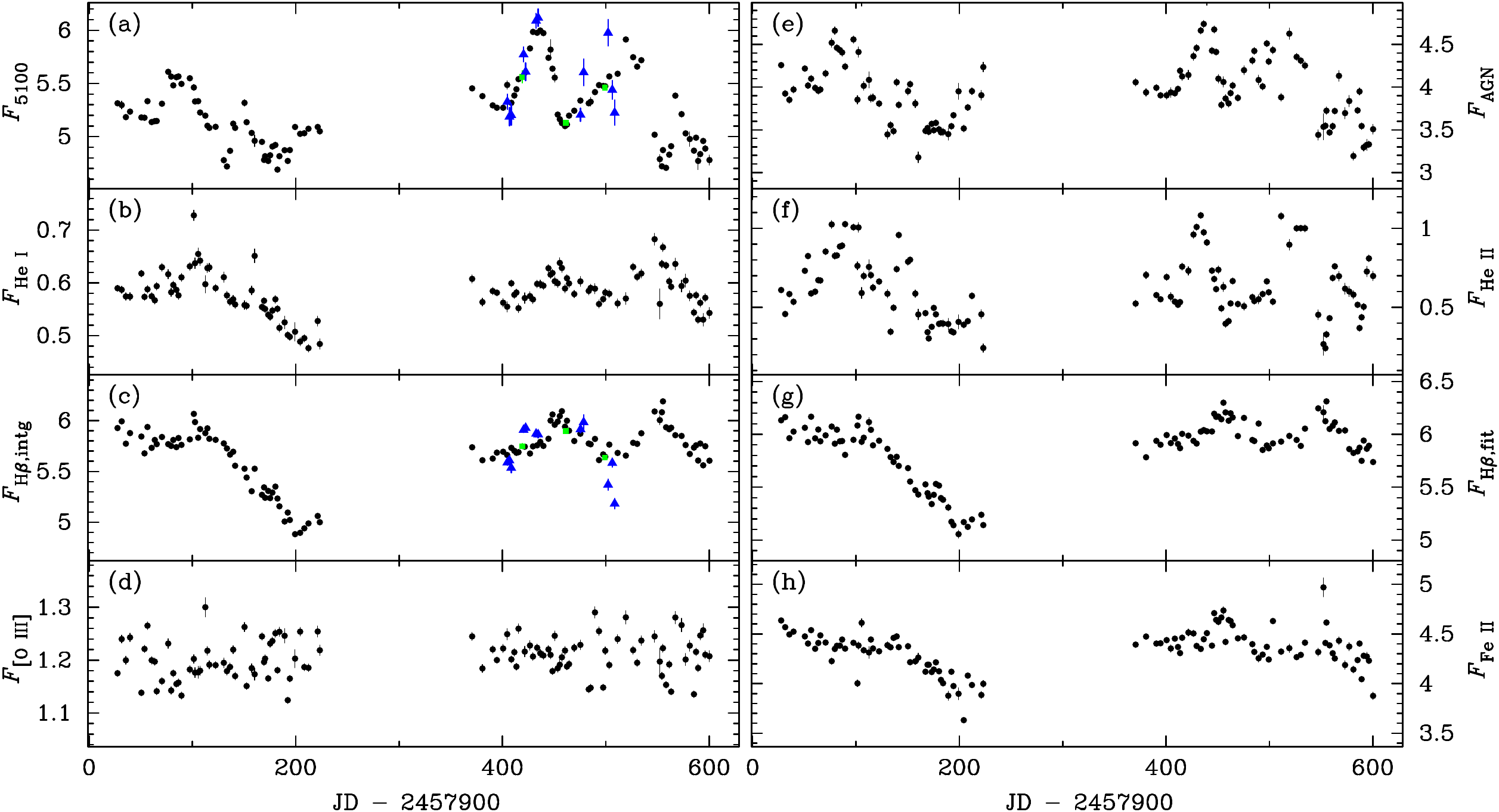}
  \caption{\footnotesize
  Light curves of continuum and emission lines. Panels (a)-(c) show
  those measured by the integration scheme, whereas (d)-(h) by the spectral
  fitting scheme. The green and blue data points in panels (a) and (c) are
  observed by Lijiang 2.4m telescope and Sutherland 1.9m telescope,
  respectively. The units of the continuum fluxes in panels (a) and (e) are
  $\times 10^{-15}\ergscma$, and the units of the emission-line fluxes in
  other panels are $\times10^{-13}\ergscm$. See details in the main text.
  }
  \label{fig-lc}
\end{sidewaysfigure*}

Figure \ref{fig-lc} panels (d)-(h) show the light curves generated
from the best-fit values of the corresponding parameters: \oiii\ flux, the
flux density of the power law at 5100 \AA\ (\fagn), \heii\ flux (\fheii), \hb\
flux (\fhbf), and \feii\ flux (\ffe). The flux of \oiii\ ideally should be a
constant \citep{peterson13}, thus the scatter of the measured \oiii\ flux can
be used to estimate the accuracy of the flux calibration. In this case, \oiii\
is relatively weak and its measurement has extra uncertainty resulting from
the blending of the \feii\ emission. So the scatter of $3\%$, which should be
considered to be an upper limit of the flux calibration accuracy, is
comparable with other RM campaigns
\citep[e.g.][]{peterson98a,kaspi00,barth15,fausnaugh17}.

The \fagn\ light curve resembles that of  \fcs, but with larger scatter. This
is caused by the uncertainty in the host galaxy component mentioned above.
The \hb\ light curves given by the fitting and simple integration are almost
the same, which is expected for a strong, single emission line \citep{hu15}.
The fluxes obtained by fitting are larger than those by integration, because
of inclusion of the flux of the broad wings which are outside of the window of
integration. 

\section{Results}

\subsection{Lags of the broad-emission lines}
\label{sec-lags}

\begin{deluxetable*}{cccccc}
  \tablewidth{10pt}
  \tablecolumns{6}
  \tabletypesize{\footnotesize}
  \tablecaption{Light curves of 5100\AA\ continuum and several lines
  \label{tab-lc}}
  \tablehead{
  \colhead{JD$-$2457900} & \colhead{\fcs} & \colhead{\fhb} & \colhead{\fhei} & 
  \colhead{\fheii} & \colhead{\ffe}
  }
  \startdata
  27.637 & 5.314$\pm$0.017 & 5.927$\pm$0.010 & 0.590$\pm$0.006 &
  0.610$\pm$0.015 & 4.636$\pm$0.020 \\
  31.593 & 5.298$\pm$0.037 & 5.993$\pm$0.012 & 0.587$\pm$0.007 &
  0.459$\pm$0.018 & 4.569$\pm$0.025
  \enddata
  \tablecomments{The 5100\AA\ continuum flux is in units of $10^{-15}{\rm
  erg\,s^{-1}\,cm^{-2}\AA^{-1}}$, and all lines are in units of
  $10^{-13}\ergs\,{\rm cm^{-2}}$. This table is available in its entirety 
  in machine-readable form.}
\end{deluxetable*}

\begin{deluxetable*}{lcccccccccccccc}
    \tabletypesize{\footnotesize}
    \tablewidth{0pt}
    \tablecolumns{15}
    \tablecaption{Reverberations of several broad-lines from the present
    campaign
    \label{tab-results}}
    \tablehead{
    \colhead{Line}           & 
    \multicolumn{2}{c}{$F_{\rm var}$}  & &
    \multicolumn{2}{c}{Lag}            & & 
    \multicolumn{2}{c}{FWHM}           & &
    \multicolumn{2}{c}{Virial Product} & &
    \multicolumn{2}{c}{CCF Broadening} \\
    \colhead{}               & 
    \multicolumn{2}{c}{(\%)}           & &
    \multicolumn{2}{c}{(days)}         & &
    \multicolumn{2}{c}{($\rm km~s^{-1}$)} & & 
    \multicolumn{2}{c}{($\times 10^7\sunm$)} & &
    \multicolumn{2}{c}{(days)} \\ \cline{2-3}\cline{5-6}\cline{8-9}\cline{11-12}\cline{14-15}
    \colhead{} & \colhead{2017} & \colhead{2018} & &\colhead{2017} &
    \colhead{2018} & &\colhead{2017} & \colhead{2018} & &\colhead{2017} &
    \colhead{2018} & &\colhead{2017} & \colhead{2018}
    }
\startdata
\heii  & 32.7$\pm$3.6 & 29.9$\pm$3.0 & &$-1.4_{-0.9}^{+6.8}$ &
$-2.9_{-0.9}^{+1.5}$ & &$6049\pm187$ & 9589$\pm$774 & &\nodata & \nodata & & 5  & $-7$ \\
\hei   & 8.2$\pm$0.9  & 4.2$\pm$0.6 & &$18.2_{-3.1}^{+7.3}$ &
$31.1_{-5.8}^{+2.9}$ & &$2547\pm18 $ & 2497$\pm$14 & &$2.30_{-0.39}^{+0.93}$ &
$3.78_{-0.70}^{+0.35}$ & & 17 & 7 \\
\hb    & 6.1$\pm$0.6  & 2.3$\pm$0.3 & &$22.6_{-3.6}^{+2.7}$ &
$27.8_{-2.9}^{+2.9}$ & &$2101\pm100$ & 2072$\pm$107 & &$1.95_{-0.36}^{+0.30}$ &
$2.33_{-0.34}^{+0.34}$ & &31 & 12 \\
\feii  & 4.8$\pm$0.5  & 3.3$\pm$0.5 & &$35.3_{-9.9}^{+8.2}$ &
$23.1_{-5.6}^{+3.4}$ & &$2047\pm94 $ & 1944$\pm$95 & &$2.88_{-0.85}^{+0.72}$ &
$1.70_{-0.45}^{+0.30}$ & &48 & 16
\enddata
\tablecomments{The virial product is defined by Equation \ref{eq:VP}.
See the text for the details of the measurements of FWHM for each line. It
should be pointed out that $F_{\rm var}$ has different meanings from the
calibration uncertainties.}
\end{deluxetable*}

We use the light curves of the continuum at 5100 \AA, \hb, and \hei\ from
integration, and \heii\ and \feii\ from spectral fitting, for the following
time-series analysis. Table \ref{tab-lc} lists all these light curves.
The errors of the fluxes listed in this table are those generated from the
errors of each pixel in the observed spectra, and are usually not large enough
to interpret the scatter in the fluxes of successive nights. So an additional
systematic error was calculated for each light curve using the same method as
in \cite{du14}, and all the analysis below was done taking account of this
systematic error.  We first calculate the variability amplitude \fvar, defined
by \citet{rodriguez97}, for each light curve. The quantity represents the
intrinsic variability over the errors in the measurements (including the
additional systematic error). Its uncertainties are calculated as defined by
\citet{edelson02}. Table \ref{tab-results} lists the results. As
mentioned in Section \ref{sec-sfs} above, the weak narrow \hb\ component has
not been subtracted from our measurements of broad \hb. The flux of \oiii\
$\lambda$5007 line is $1.21\pm0.04\times10^{-13} \ergscm$, measured from the
light curve of \oiii\ in panel (d) of Figure \ref{fig-lc}. Assuming a typical
value of 0.1 for the narrow \hb/\oiii\ intensity ratio in AGNs
\citep[e.g.,][]{veilleux87}, the narrow \hb\ component contributes only
$\sim$2\% fluxes of the \hb\ measured here. Thus, the dilution of \hb\ \fvar\
by the uncorrected narrow component is negligible. Note that \heii\ has an
\fvar\ much larger than the other emission lines, and even the continuum. This
behavior is also shown by the strong \heii\ emission feature in the RMS
spectrum (see the bottom panel of Fig. \ref{fig-rms}), and common for many
objects in previous RM investigations \citep{barth15}.

\begin{figure*}
  \centering
  \includegraphics[width=0.4\textwidth]{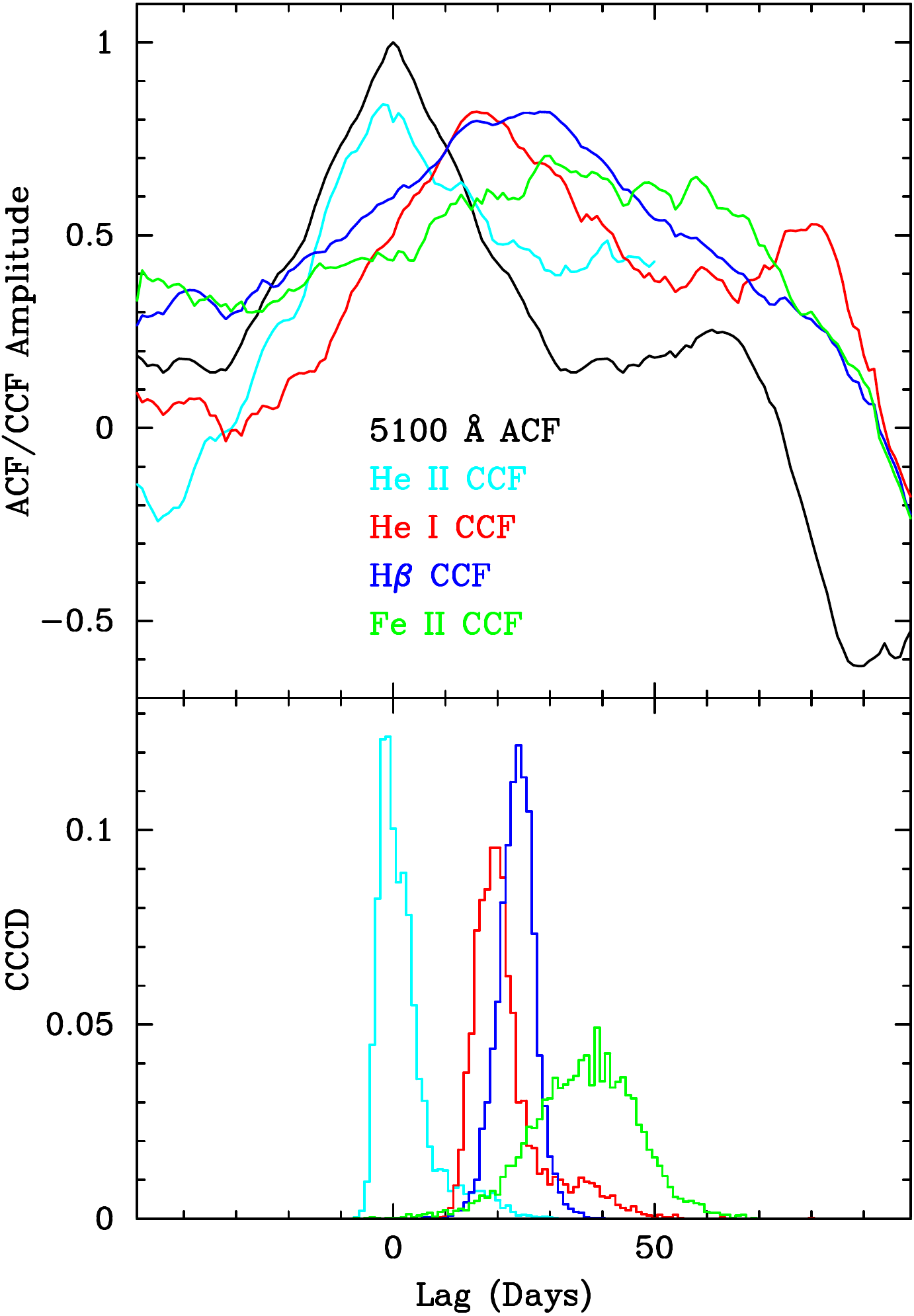}
  \includegraphics[width=0.4\textwidth]{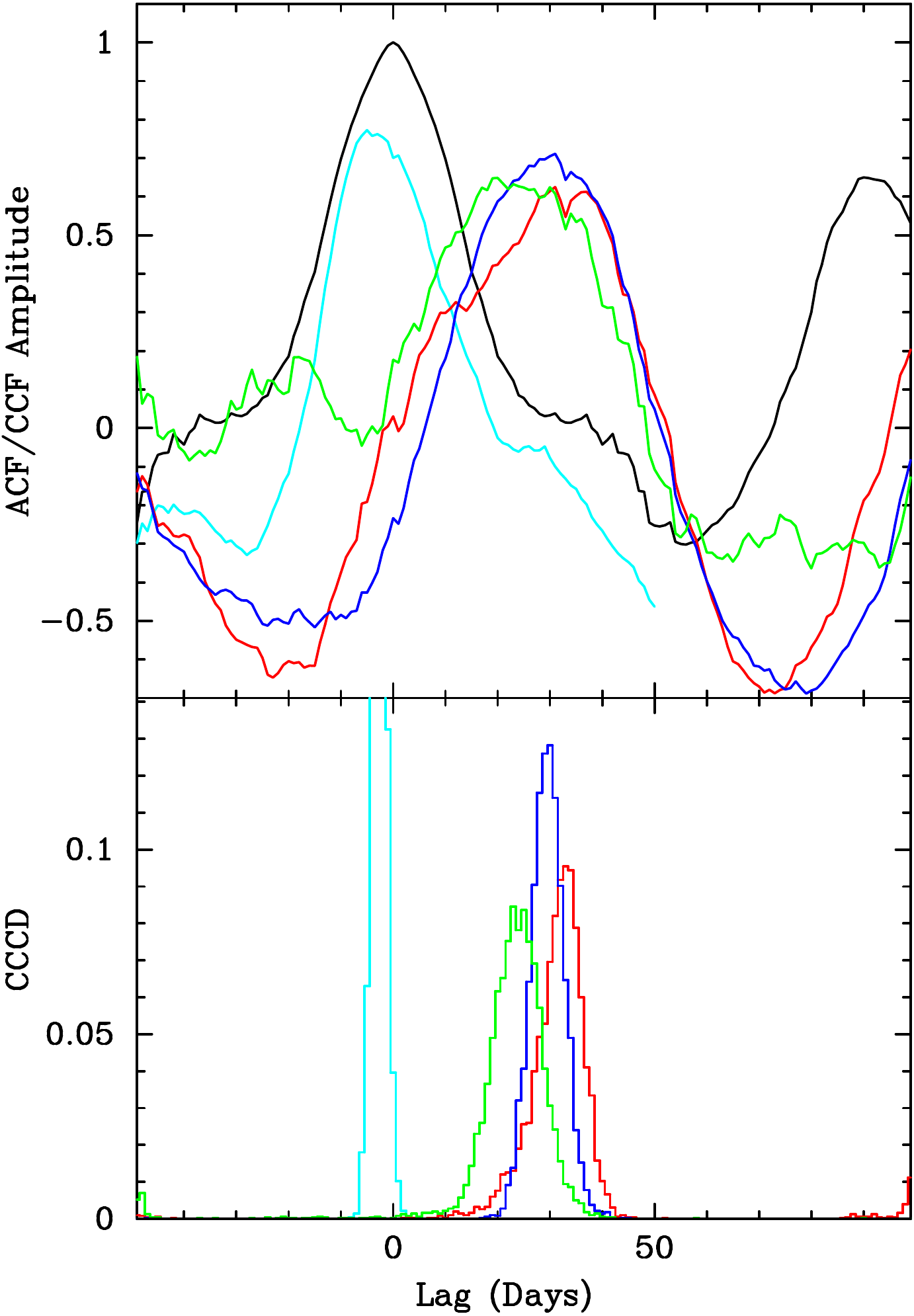}
  \caption{\footnotesize Cross-correlation analyses of H$\beta$, \hei,
  \heii\, and \feii\, lines with 5100\AA\, continuum, for 2017 (left)
  and 2018 (right), respectively.
  }
  \label{fig-ccf}
\end{figure*}

We measured the reverberation lags between the variations of the continuum
(\fcs) and emission lines (\fheii, \fheis, \fhbs, and \ffe), using the
standard interpolation cross-correlation function (CCF) method
\citep{gaskell86,gaskell87,white94}. The centroid of the CCF, using only
the part above 80\% of its peak value (\rmax), is adopted as the time lag
\citep{koratkar91,peterson04}. The error of the time lag is given by the
15.87\% and 84.13\% quantiles of the cross-correlation centroid distribution
(CCCD) generated by 5000 Monte Carlo realizations, following
\citet{maoz89,peterson98b}. In each realization, a subset of data points is
selected randomly, and their fluxes are also changed by random Gaussian
deviations according to the errors. The top panel of Figure \ref{fig-ccf} shows
the autocorrelation function (ACF) of the \fcs\ light curve (in black), and
also the CCFs for the emission lines (\heii\ in cyan, \hei\ in red, \hb\ in
green, and \feii\ in blue) with respect to \fcs. The bottom panel shows the
corresponding CCCDs for the CCFs. The measured time lags are listed in 
Table \ref{tab-results}.

\subsection{Line-profile measurements}

For \hb\ and \feii, which are included in the fitting and allowed to vary, the
means of their velocity widths and shifts obtained from the best fits to
individual-night spectra are adopted to be the measurements of their line
profiles. The standard deviations are used as the errors. For \heii, the
velocity width and shift are measured from the rms spectrum. 
For \hei, we perform an additional fit to a narrow band of the spectrum
around the emission line including only a straight line as the continuum
and a single Gaussian for \hei. Their errors are estimated by Monte
Carlo simulation: several realizations of mean and rms spectra were generated
by bootstrap sample selection, and the same spectral fitting was performed and
the velocity widths were measured. Then the standard deviations of the
velocity widths measured from the realizations are used as the errors.

\begin{deluxetable*}{lcccccccc}
\renewcommand{\arraystretch}{1.5}
    \tablecaption{H$\beta$ line width information\label{tab:sigma}}
    \tablecolumns{7}
    \tabletypesize{\footnotesize}
    \tablewidth{0pt}
    \tablehead{
        \colhead{}                        &
        \multicolumn{2}{c}{mean spectra}          & 
        \colhead{}                        &
        \multicolumn{5}{c}{rms}                   \\ \cline{4-9}
        \colhead{}                        &
	\multicolumn{2}{c}{$\sigma_{\rm line}$}     &
        \colhead{}                        &
	\multicolumn{2}{c}{FWHM}                    &
        \colhead{}                        &
	\multicolumn{2}{c}{$\sigma_{\rm line}$}   \\ 
	\cline{2-3} \cline{5-6} \cline{8-9}
	\colhead{Year} & \colhead{2017} & \colhead{2018} &
        \colhead{}                        &
	\colhead{2017} & \colhead{2018} & 
        \colhead{}                        &
	\colhead{2017} & \colhead{2018}
            }
\startdata
Width & $1437\pm64$ & $1438\pm41$ & & $2447\pm73$ & $2054\pm557$ & & $1272\pm97$ & $874\pm237$ \\
$\hat{M}_{\bullet}(10^7\sunm)$ & $0.91_{-0.17}^{+0.14}$ &
$1.12_{-0.13}^{+0.13}$ & & $2.64_{-0.45}^{+0.35}$ & $2.29_{-1.26}^{+1.26}$ & &
$0.71_{-0.16}^{+0.14}$ & $0.41_{-0.23}^{+0.23}$
\enddata
\tablecomments{\footnotesize  The width is in units of \kms\, and after
instrumental broadening correction. Here the values were measured from the rms
spectra shown in Figure \ref{fig-rms}, without the deconvolution in Section
4.3. The blue peak around $-2500$ \kms\ in the rms spectrum after
deconvolution in Figure \ref{velocity-resolved} is not high enough to
contribute the FWHM listed here.}
\end{deluxetable*}

The resolution is estimated by comparing the width of \oiii\ in our spectra
(FWHM = 1084 \kms) to those from previous high-spectral-resolution
observations.  \cite{Whittle1992} obtained an \oiii\ FWHM measurement of 350
\kms\ (see also \citealt{grier12}), yielding a broadening of 1020 \kms. We also
fit the spectrum of this object from \cite{bg92}, obtained an \oiii\ FWHM of
400 \kms\ after correcting for their instrument broadening. The resultant
broadening is similar, and we adopted a FWHM = 1000\kms\ as the instrument
broadening of our spectra. The measurements of the widths of each
emission line after instrument broadening correction are listed in Tables
\ref{tab-results} and \ref{tab:sigma}.

\subsection{Velocity-resolved delays}

In order to investigate the geometry and kinematics of the BLR in PG 2130+099,
we calculated the velocity-resolved time lags of the \hb\ emission line
profile. First, the influence of the varying line-broadening functions
($\psi$) and wavelength calibration inaccuracy for different nights, which are
caused by the changing seeing and misalignment of the slit, must be removed.
Using comparison stars, the line-broadening function can be obtained by
fitting the spectra of the stars using a stellar template convolved by $\psi$.
The wavelength calibration inaccuracy is also taken into account as the
velocity shift of $\psi$. We adopted the Richardson-Lucy deconvolution
algorithm, which is demonstrated to be an efficient way to recover the signal
blurred by a known response kernel, to correct the $\psi$ function. Details of
these mathematical manipulations are described by \citet{Du2016a}. After the
correction, the emission line is divided into several bins, each of which has
the same flux in the rms spectrum. Then, the light curves in the bins and
their corresponding time lags relative to the continuum light curve are
measured as previously described.

\begin{figure*}
  \centering
  \includegraphics[width=0.42\textwidth]{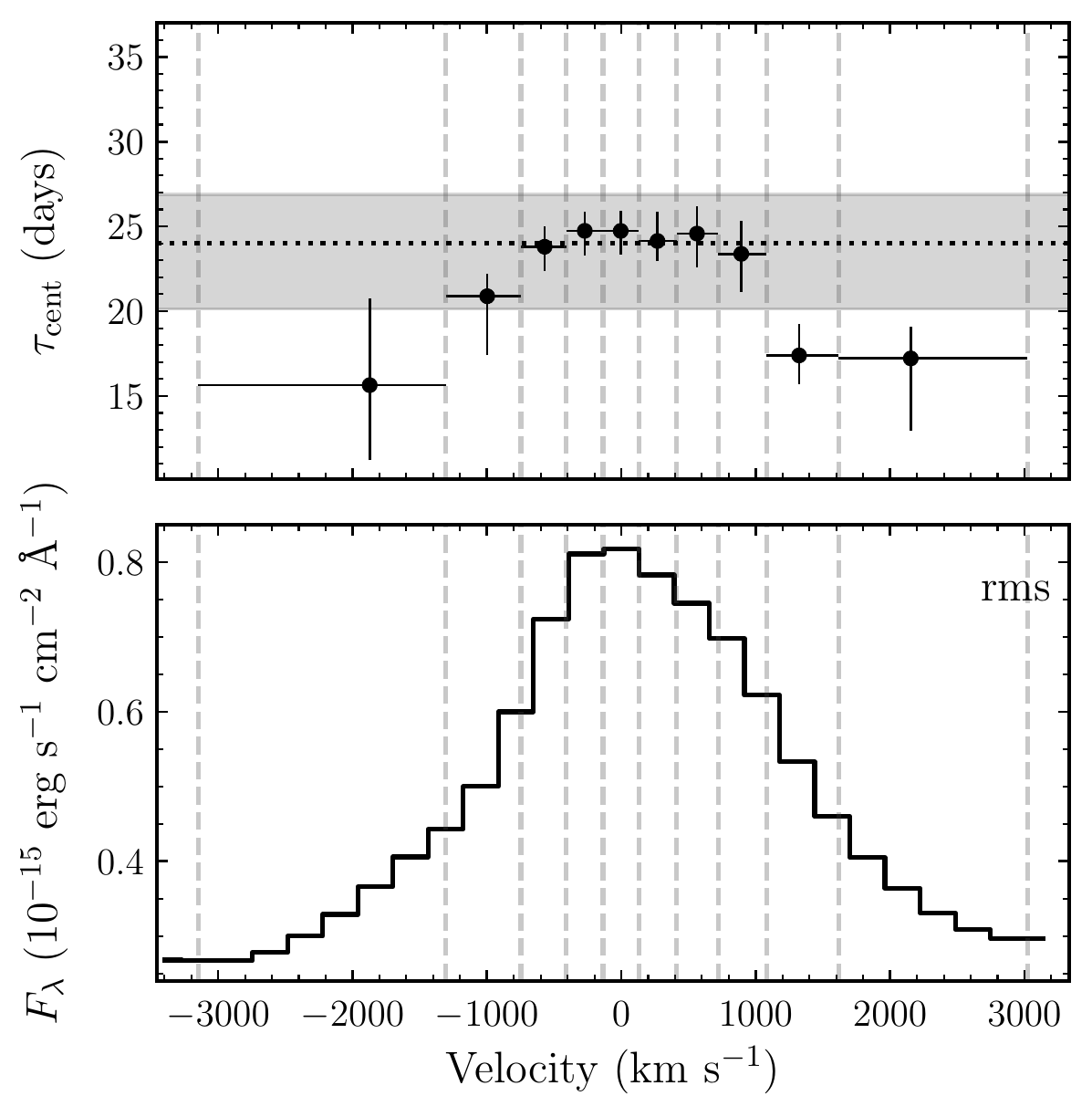}
  \includegraphics[width=0.42\textwidth]{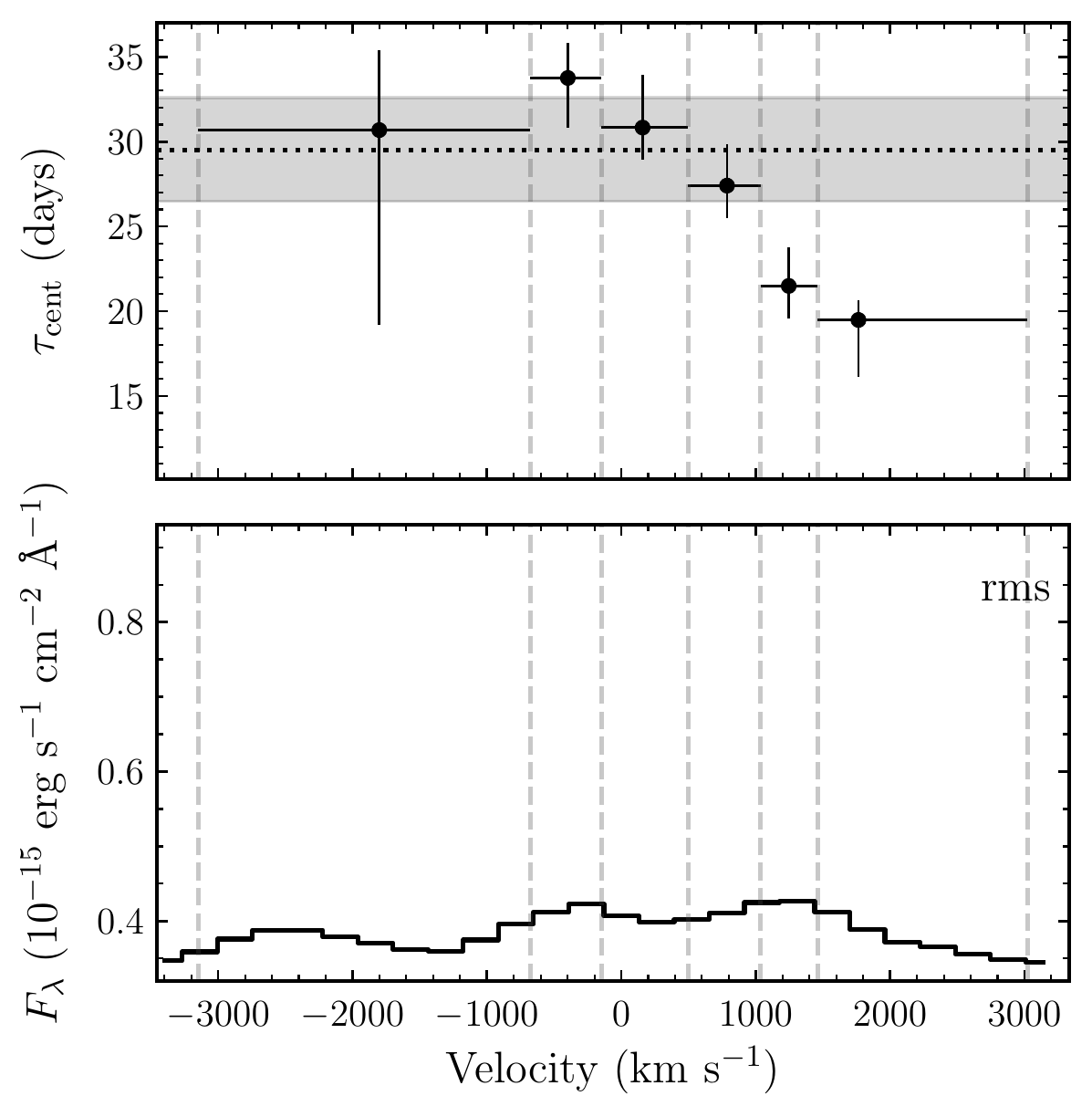}
  \caption{\footnotesize Velocity-resolved delays from 2017 (left) and
  2018 (right) observations, showing very different shapes. Observations in
  2017 show shorter time delays for high-velocity gas, consistent with the
  model of the \hb-emitting region being in a virialized motion. In
  contrast, observations in 2018 show shorter lags in the red wing, which is
  more likely generated from an infall. Light curves and CCF analysis are
  given for each velocity bin in Appendix A.
  }
  \label{velocity-resolved}
\end{figure*}

Figure \ref{velocity-resolved} shows the velocity-resolved time delays and the
rms spectrum (after the line-broadening correction) obtained for 2017
(left) and 2018 (right), respectively. The figure indicates that the BLR
in 2017 has a geometry of virialized motion (e.g., see plots in
\citealt{Welsh1991},\citealt{Bentz2009}), while in 2018 it more likely
has an infall structure. However, it should be pointed out that the bluest
point in the right panel of Figure \ref{velocity-resolved} has quite large
error bars because of the weak response in the second year. Whether or not the
BLR kinematics are consistent with inflow remains open, but it is certain that
the variable BLR has changed between the two years. For the upcoming
discussion, we refer to the BLR as an inflow in the second year. The exact
description of the geometry should be given by the application of the
maximum-entropy method (MEM) developed by \cite{horne04}, which is out of the
scope of this paper.

We note that the lags in both wings could be contaminated by the continuum
subtraction. In order to test the contamination, we show the velocity-resolved
delays with equal fluxes in the mean spectra in Appendix A. Light curves of
each velocity bins are shown in Appendix B as well as CCF analysis. Comparing
them, we conclude that the influence is quite weak.

The data quality of other emission lines is not good enough to determine the
velocity-resolved time lags.

\subsection{Transfer functions}

Following the procedure 
in \citet{Li2016}\footnote{The procedure is implemented in the package
\text{MICA}, which is publicly available at
\url{https://github.com/LiyrAstroph/MICA2}}, we used a Gaussian to
parameterize the transfer function of each emission line. The amplitude,
width, and central values of the Gaussian are free parameters and determined
by a Markov-chain Monte Carlo method. The continuum variations are described
by a damped random walk model. As such, the co-variance functions between
continuum and emission lines and between lines can be expressed analytically.
This allowed us to employ the well-established framework given by
\citet{Rybicki1992} to calculate the Bayesian posterior probabilities and infer
the best values for free parameters.

\begin{figure*}
  \centering
  \includegraphics[width=0.48\textwidth]{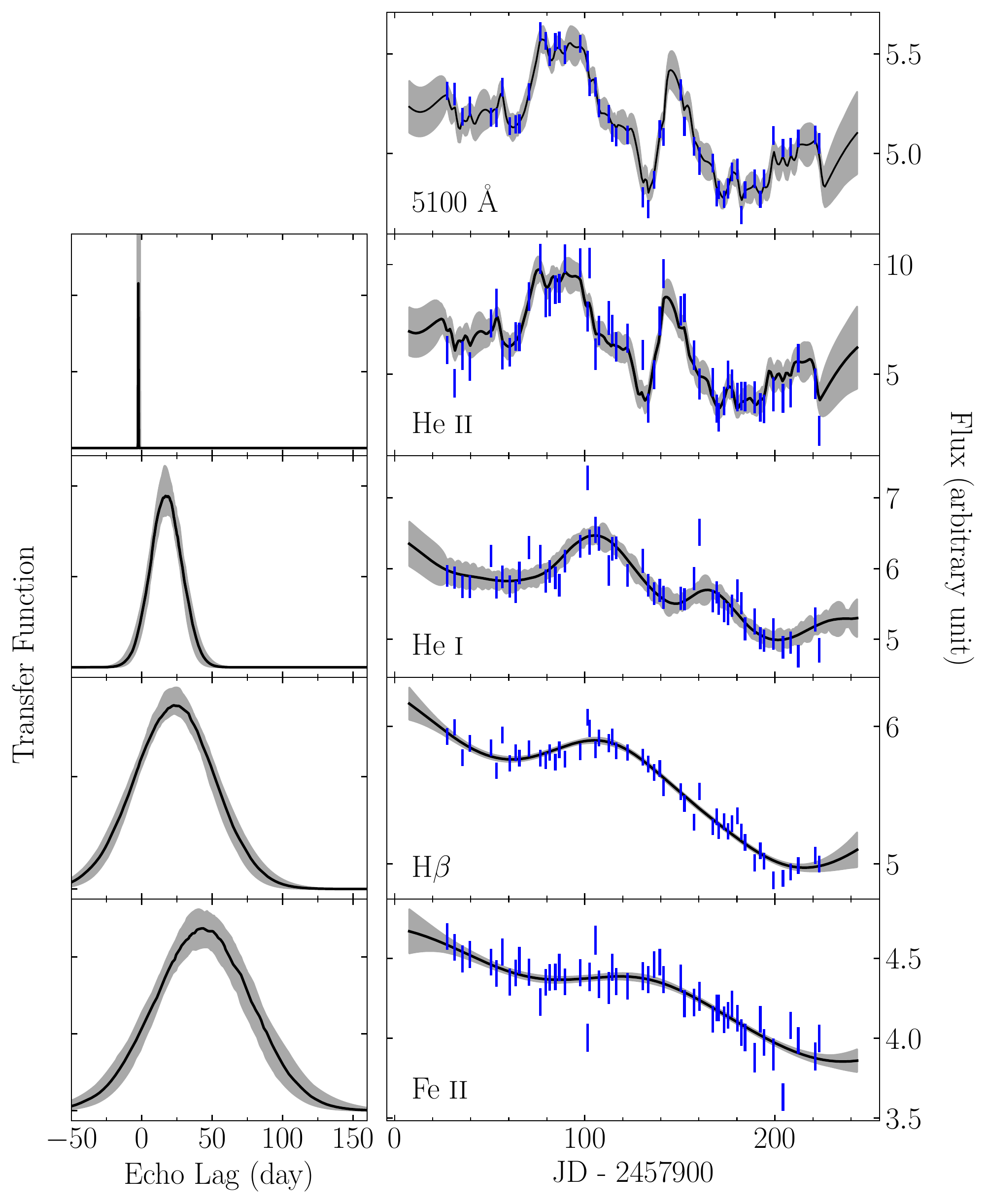}
  \includegraphics[width=0.48\textwidth]{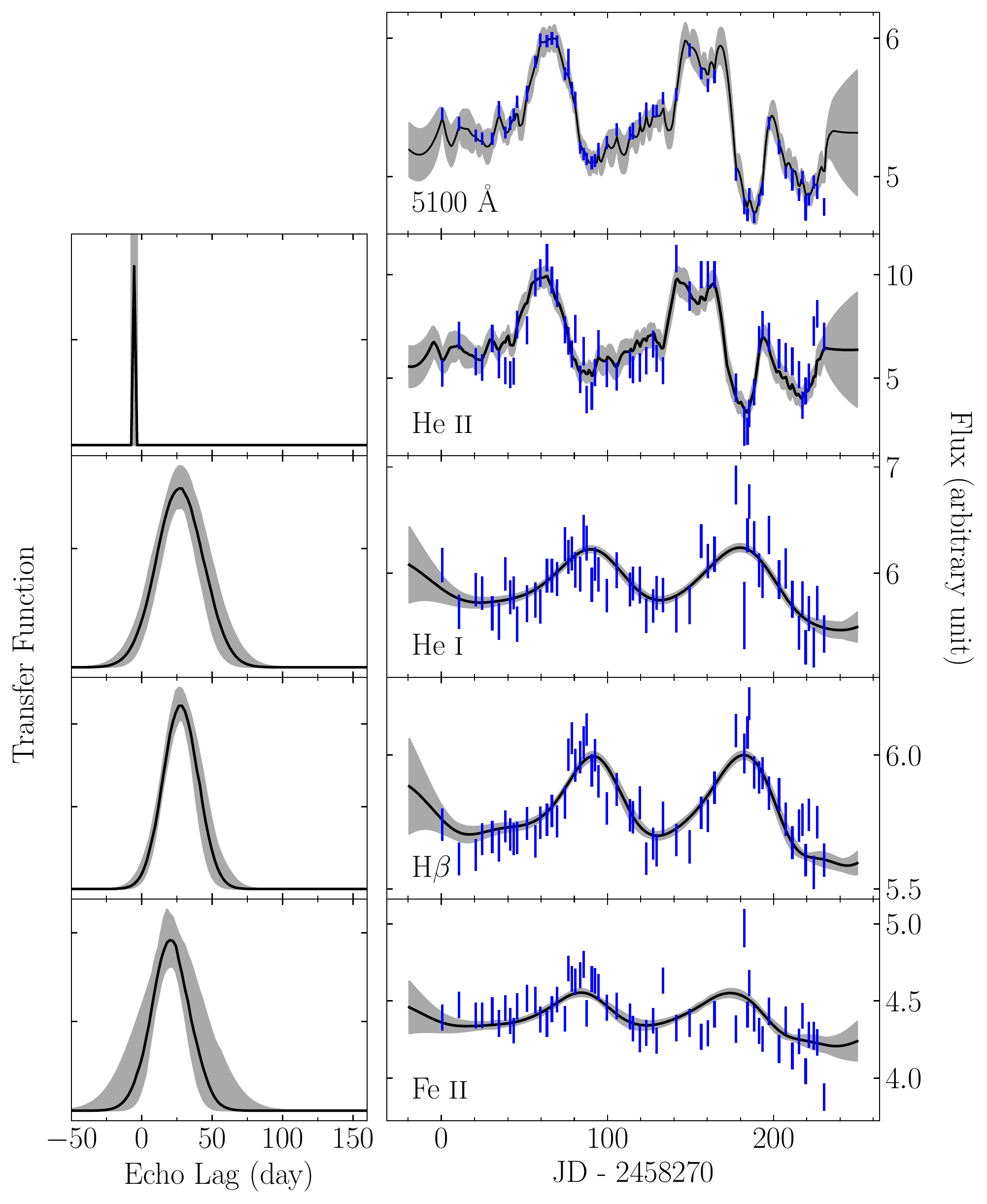}
  \caption{\footnotesize Transfer functions of the \hb, \hei, and \feii\ lines
  from the first (left) and the second year (right).  }
  \label{fig:broadening}
\end{figure*}

\begin{deluxetable*}{cccccccccccc}
\renewcommand{\arraystretch}{1.2}
    \tablecaption{BLR lags and broadening from the transfer functions\label{tab:broadening}}
    \tablecolumns{12}
    \tabletypesize{\footnotesize}
    \tablewidth{0pt}
    \tablehead{
        \colhead{Campaign}      &
        \mcn{3}{c}{\hb}         &
        \colhead{}              &  
        \mcn{3}{c}{\hei}        &
        \colhead{}              &
        \mcn{3}{c}{\feii}       \\   \cline{2-4} \cline{6-8}\cline{10-12} 
     & $\tau$& $\Delta \tau$&$\tau/\Delta\tau$& &$\tau$&$\Delta \tau$&$\tau/\Delta\tau$& 
&$\tau$&$\Delta \tau$ & $\tau/\Delta\tau$
}
\startdata
2017 & $23.2\pm2.3$ & $28.3\pm3.2$& $0.8\pm0.1$& &$17.1\pm1.6$& $11.0\pm1.8$&$1.6\pm0.2$& 
&$43.3\pm4.9$& $34.9\pm4.2$ &$1.3\pm0.2$\\
2018 & $27.6\pm1.9$& $12.5\pm2.6$ &$2.2\pm0.3$& &$27.3\pm2.1$ &$16.9\pm3.1$ &$1.6\pm0.3$& 
&$20.5\pm2.1$&$14.1\pm7.1$&$1.4\pm0.6$
\enddata
\tablecomments{$(\tau,\Delta\tau)$ are in units of days.}
\label{tab:TF}
\end{deluxetable*}

Fitting results for the two years are shown by Figure \ref{fig:broadening} and
listed in Table \ref{tab:broadening}, separately. Lags obtained by the
transfer functions are consistent with those of the CCF method.

\subsection{Black hole mass and accretion rates}
\label{BHmass}

The bulge velocity dispersion of PG 2130+099 is $\sigma_*=163\pm 19\,$\kms
from near-infrared spectra \citep{Grier2013a}, but the host galaxy has a
pseudo-bulge ($n\approx 0.45$) and a disturbed disk \citep{Kim2017}. Using the
same slope of the $\mbh-\sigma_*$ relation \citep{kormendy13} but the half of
its zeropoint \citep{Ho2015}, we expect $\mbh\approx 6.5\times 10^7\sunm$ in
PG 2130+099.  This estimate should be regarded as an upper-limit. Indeed,
\cite{Grier2017} made use of a Markov Chain Monte Carlo (MCMC) method to model
the BLR and to determine the black hole mass through H$\beta$ light curves,
and found ${\rm log_{10}}(\mbh/\sunm)=6.87_{-0.23}^{+0.24}$, which is
significantly lower than that from the $\mbh-\sigma$ relation.
Comparing our velocity-resolved time lags in 2017 as shown by Figure
\ref{velocity-resolved} with the MCMC model \citep{Grier2017}, we find the
model matches our measurements, leading to support of
$\mbh=10^{6.87}\sunm$ as a conservative mass estimation.

Using the presently detected \hb\, lags, we determine the virial product 
\begin{equation}
\hat{M}_{\bullet}=\frac{c \tauhb  \times V^2}{G},
\label{eq:VP}
\end{equation}
where $G$ is the gravitational constant, $V$ is either full-width-half-maximum
($V_{\rm FWHM}$) or velocity dispersion ($\sigma_{\rm line}$) of the \hb\
profile in the mean spectrum or rms. We list $\hat{M}_{\bullet}$ in Table
\ref{tab-results} and \ref{tab:sigma}, yielding the black hole mass if given the
virial factor ($\fblr$). The $\fblr$ can in principle be calibrated by the
$\mbh-\sigma_*$ relation (e.g., \citealt{onken04,Woo2015,Battiste2017}), or
the black hole masses from accretion disk models \citep{Lv2008,Mejia2017}, but
it tends to be smaller for AGNs with pseudo-bulges, such as $\fblr=0.5$
\citep{Ho2014}. If we take $\fblr=0.5$, we have
$\mbh=0.97_{-0.18}^{+0.15}\times 10^7~\sunm$ from the virial product by using
the mean FWHM of \hb\ in 2017 (the BLR may significantly deviate from
virialized state in the second year), agreeing with the MCMC result of black
hole mass in \citep{Grier2017}. 

In light of the standard model of accretion disks \citep{Shakura1973}, the
dimensionless accretion rates defined by $\mathdotM=\dot{M}_{\bullet}/L_{\rm
Edd}c^{-2}$ can be estimated as \citep{du16b}
\begin{equation}
\mathdotM= 20.1\left(\frac{\ell_{44}}{\cos i}\right)^{3/2}M_7^{-2},
\end{equation}
if given the optical luminosity and the black hole mass, where
$\dot{M}_{\bullet}$ is the accretion rate,  and $L_{\rm Edd}$ is the 
Eddington luminosity, $\ell_{44}=L_{5100}/10^{44}\ergs$, $M_7=\mbh/10^7\sunm$,
and $i$ is the inclination of the disks (we take $\cos i=0.75$
\footnote{%
\citet{Grier2017} gives a value of $30.2_{-10.1}^{+11.0}$ degrees for the
inclination angle of this object from dynamically modeling, which is
marginally consistent with our choice here. We still take $\cos i=0.75$ for a
convenient comparison with other objects without inclination angle
measurements.
}%
). The mean flux at 5100\AA\, in the first year is
$\bar{F}_{5100}=(5.12\pm0.25)\times 10^{-15}{\rm
erg\,s^{-1}\,cm^{-2}\,\AA^{-1}}$ corresponding to a mean luminosity $\lambda
L_{\lambda}=(2.50\pm0.12)\times 10^{44}\ergs$ in the present epoch. With
$\mbh$ and $\ell_{44}=2.50$, we obtain $\mathdotM=10^{2.1\pm0.5}$, implying a
super-Eddington accretor.

\section{Discussions: unresolved questions}

Results from the two years are shown in the previous section, but several
puzzles arise. From Table \ref{tab-results}, we find $R_{\rm H\beta}^{\rm
1st}=22.6_{-3.6}^{+2.7}$ ltd and $R_{\rm H\beta}^{\rm 2nd}=27.8_{-2.9}^{+2.9}$
ltd, showing that the \hb\ region slightly changed in the two years
(but the difference is less than two times of the uncertainty, so
could be caused by just uncertainties in the measurements). However, the
radii of the \hei\ and \feii\ regions changed by a factor of almost 2.

It is useful to illustrate the dynamical timescale of the BLR variations
defined as
\begin{equation}\label{eq:tBLR}
\Delta t_{\rm BLR}=\frac{R_{\rm H\beta}}{V_{\rm H\beta}}=8.2\,R_{20}V_{2000}^{-1}\,{\rm yr},
\end{equation}
where $R_{20}=R_{\rm H\beta}/20$ ltd is the BLR size in units of 20\,ltd and
$V_{2000}=V_{\rm H\beta}/2000$ \kms\ is the FWHM(\hb) in units of 2000 \kms.
It implies that the structure and kinematics of the BLR in PG 2130+099 cannot
be significantly changed in two successive years but either the ionization
structure could follow the variations of the ionizing source in one year, or
some physical processes drive internal changes (i.e., some local variations in
the BLR).

Note that the time lag of \heii\, we measured is negative, but consistent with
zero considering its uncertainty. Similar results have been reported for other
objects (e.g. \citealt{barth13}). Keeping in mind the result that \heii\ has a
much larger variability amplitude than $F_{5100}$, a possible explanation of
the negative/zero time lag is that \heii\ is responding to the UV ionizing
flux, which leads the optical continuum, as suggested by, e.g., \cite{barth13}
and \cite{Edelson2019}.

\subsection{Onion structure}

\begin{figure}
  \centering
  \includegraphics[width=0.45\textwidth]{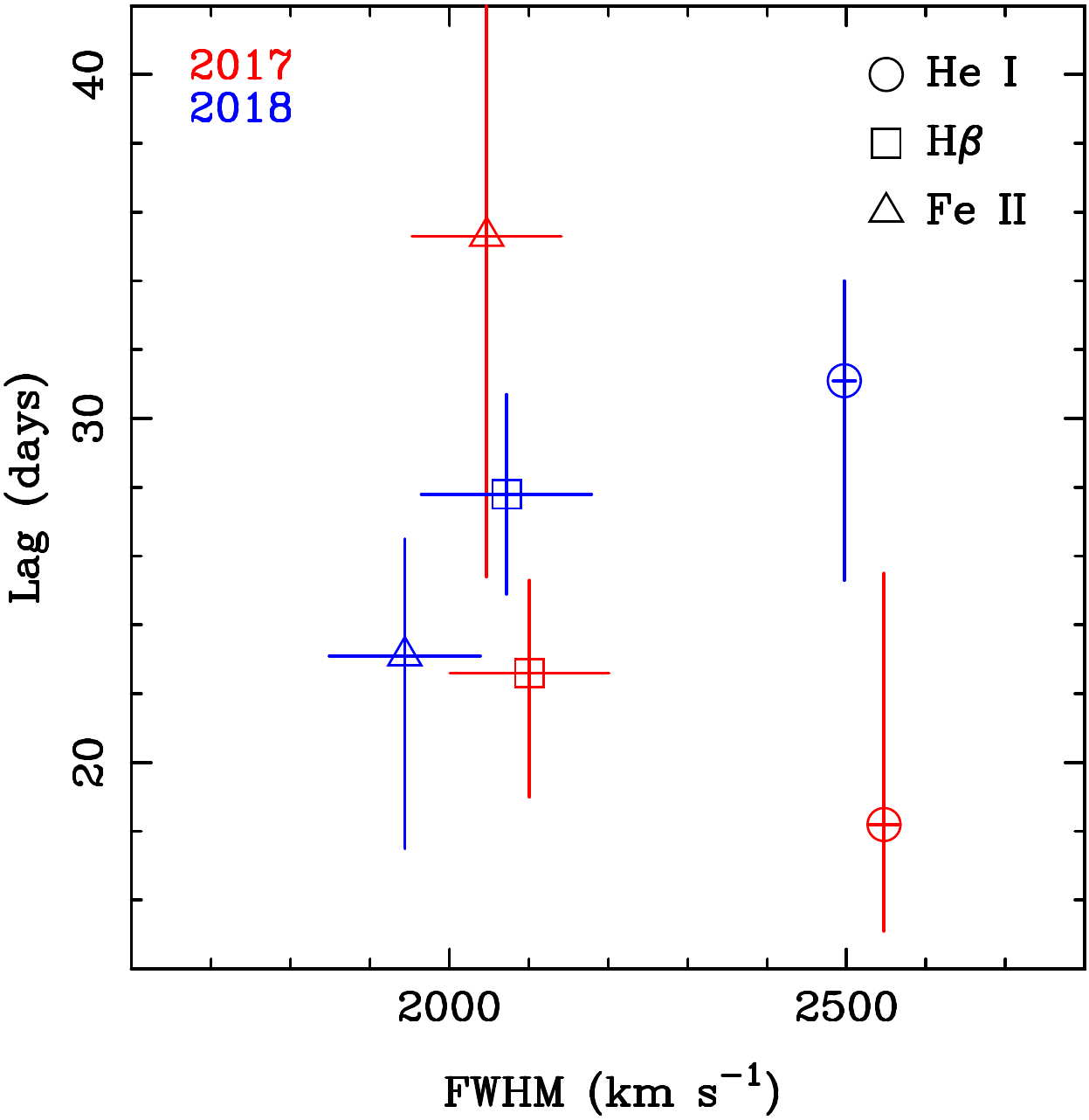}
  \caption{Time lag vs. FWHM of emission lines. The BLR does not
  conform to the onion structure in 2018.}
  \label{fig-lagwidth}
\end{figure}

The onion structure of the BLR is quite well understood as a sequence of
ionization energy of ions \citep{collin88} and FWHM of lines
\citep{peterson99}.
The ions relevant to this study have ionization energies of $\epsilon_{\rm
He~II}=54.4$ eV, $\epsilon_{\rm He~I}=24.58$ eV, $\epsilon_{\rm H\beta}=13.6$
eV and $\epsilon_{\rm Fe~II}=7.87$ eV. From the first year, we have $R_{\rm
Fe~II}>R_{\rm H\beta}>R_{\rm He~I}>R_{\rm He~II}$, following the
onion structure as shown by Figure \ref{fig-lagwidth}. 

In the observations of the second year, \hb-emitting region do not show
significant variations, however, \hei\, and \feii\, regions show significant
changes, in particular $R_{\rm Fe~II}\lesssim R_{\rm H\beta}\lesssim R_{\rm
He~I}$ disrupting the onion structure with a timescale much less than the
dynamical time as shown by Equation (\ref{eq:tBLR}). The first unresolved
question would be to explain the mechanism that generates such a change
in the onion structure.

\subsection{Kinematics}

The velocity-resolved delays of the broad \hb\ line in the two years shown in
Figure \ref{velocity-resolved} are different. The responding emission line
region show a change from a virialized motion to an inflow on a
timescale of less than one year. The profile and the intensity of the \hb\
line in the rms spectra dramatically change. It is a smooth core-dominated
profile in the first year, but multiple-peaked and wing-dominated in the
second. 
On the other hand, the widths
of the broad lines in the mean spectrum remain almost the same as in the first
year, as well as the virial products derived.
This implies that the BLR may still be actually virialized in the
second year. Such an inconsistency worths further investigations: is the
velocity-resolved delays in the second year a unique signature of infall as in
those simple models? If so,
why the majority of \hb-emitting clouds have virialized motions in the
second year as shown by the mean spectrum, while the reverberation part has a
kinematics of inflow indicated by the velocity-resolved delays? Further more,
what is the connection between the inflow and the virialized part of
the BLR? This is the second puzzle, which motivates us to continue monitoring
PG2130+099.

\subsection{Weak reverberation}

The variability amplitude of 5100\AA\, flux ($F_{\rm var}=6.7\%$) in the second
year is larger than that in the first year ($F_{\rm var}=4.7\%$), however,
H$\beta$ is less variable in the second year ($F_{\rm var}=2.3\%$) than in the
first year ($F_{\rm var}=6.1\%$). This means a weaker response of \hb\ to the
varying continuum in the second year. That the virialized part of the
BLR did not respond in the second year may be consistent with the weak
response, but it is hard to understand why. The transfer functions are also
very different in the two years. Though the mean radius of \hb-emitting region
does not change much, the width ($\tau/\Delta \tau$) changes by a factor of 2
(see Table \ref{tab:TF} and Figure \ref{fig:broadening}). On the contrary,
\hei\, and \feii\, mean radii change significantly, but their relative widths
($\tau/\Delta\tau$) stay almost unchanged. This is the third puzzle.

Photoionization calculations and light curves simulations have been preformed
in the literature \citep[e.g.][]{korista04,goad14,lawther18} to investigate
how the emission-line responsivity depends on factors, including the continuum
state, driving continuum variability, BLR geometry, and the duration/cadence
of the campaign. \citet{korista04} calculated photoionization models suitable
for NGC 5548, and found that the local line responsivity could be two times
lower in the high continuum state than that in the low state (their Figure 3)
with a factor of $\sim$8 times change in the continuum flux. In this case,
the \hb\ responsivities, estimated using the \fvar\ ratio of the line to the
continuum as suggested in \citet{goad14}, are $\sim$1.3 and 0.34 for the two
years, respectively. But the continuum flux $F_{5100}$ is only slightly higher
($<$5\% on average, see Figure \ref{fig-lc} panel a) in the second year.
Note that the ionizing UV continuum may has larger variability between
the two years than $F_{5100}$. Its amplitude and impact on the responsivity
need to be confirmed by further observations and calculations.
\citet{goad14} shown that continuum variations faster than the maximum time lag
for an extended BLR reduce the measured line responsivity and also the time
lag in the meantime (their Figure 9). The continuum variability time scale in
the second year seems to be somewhat faster than that in the first year by
comparing the widths of the two ACFs in Figure \ref{fig-ccf}, but the measured
\hb\ time lag does not become shorter accordingly as expected if the
variability time scale is the reason for the weaker response. Thus, factors
other than the continuum flux state and variability time scale are needed to
interpret the much weaker resposivity in the second year. Detailed
calculations and simulations out of the scope of this paper are necessary to
answer this question.

\subsection{Comparing with previous campaigns}

As mentioned previously, three campaigns have been done for PG 2130+099,
including \cite{kaspi00,grier08,grier12} before the present work. Since the
\citet{kaspi00} campaign has poor sampling and the \citet{grier08} campaign is
too short in duration, 
we only compare our results with \citet{grier12}, which performed a cadence of
roughly one day. \citet{grier12} obtained $\tau_{\rm H\beta}=9.7\pm 1.3$ days,
which is shorter than the value measured in this campaign by a factor of
larger than two. Considering that our campaign is 7 years later after that of
\citet{grier12}, the difference could be real and due to changes of the BLR.
However, \cite{Bentz2013} argued that the \citet{grier12} campaign missed some
key observation dates and PG 2130+099 should have a lag of $31\pm4$ days from
their reanalysis. This corrected time lag is in agreement with the present
results.

\section{Summary}

We report a successive two-year campaign of PG 2130+099 using the CAHA
2.2m and other telescopes since 2017. Reverberations of several broad
emission lines are analysed. We find that
\begin{itemize}
  \item From the observations in 2017, \hb\ displayed a lag of
    $\tau_{\rm H\beta}=22.6_{-3.6}^{+2.7}$ days.
    \hei, \heii\, and \feii\ have lags of $\tau_{\rm He~I}=18.2_{-3.1}^{+7.3}$
    days, $\tau_{\rm He~II}=-1.4_{-0.9}^{+6.8}$ days, and $\tau_{\rm
    Fe~II}=35.3_{-9.9}^{+8.2}$ days, respectively. The velocity-resolved
    delays of \hb\ shows a sign of virialized motion. The BLRs are
    radially stratified (i.e., an onion structure) according to the
    full-width-half-maximum of the lines and the ionization energies of the
    ions.
  \item From the observations in 2018, we obtain $\tau_{\rm
    H\beta}=27.8_{-2.9}^{+2.9}$ days, $\tau_{\rm He~I}=31.1_{-5.8}^{+2.9}$
    days, $\tau_{\rm He~II}=-2.9_{-0.9}^{+1.5}$ days,  and $\tau_{\rm
    Fe~II}=23.1_{-5.6}^{+3.4}$ days. The \hb\ velocity-resolved delays favour an
    inflow as the variable part of the BLR. It is clear that the onion
    structure is broken in 2018 with a timescale less than one year.
  \item We prefer the estimation of black hole mass from the first year
    observations. Using the black hole mass of
    $\mbh=0.97_{-0.18}^{+0.15}\times 10^7~\sunm$, we estimate an
    accretion rate of $10^{2.1\pm0.5}~L_{\rm Edd}/c^2$, suggesting that it is a
    super-Eddington accretor.
  \item Our two-year high-cadence campaign shows several puzzles of the BLR
    reverberations in PG 2130+099. The stratified structure of \hb, \hei, and
    \feii-emitting regions changes, and the kinematics of \hb-emitting region
    changes from a virialized motion to an inflow, with a timescale
    less than one year. The mean spectra are less variable, but the line
    responsivity becomes much weak in the second year. It is worth continuing
    intensive monitoring of PG 2130+099 to resolve these questions.
\end{itemize}

PG 2130+099 shows interesting reverberation behavior in our high-cadence
campaign. Markov Chain Monte-Carlo simulations of the BLR
\citep{pancoast11,Li2018} will be carried out for detail explanations and
better understandings.
       
\acknowledgments{The authors are grateful to an anonymous referee for very
useful reports. The campaign made use of mainly the CAHA 2.2m, and jointly the
Lijiang 2.4m and the Sutherland 1.9m telescopes, we acknowledge the support of
the staffs of these telescopes. Funding for the Lijiang 2.4m telescope has
been provided by Chinese Academy of Sciences (CAS) and the People's Government
of Yunnan Province. This research is supported by grant 2016YFA0400700 from
the Ministry of Science and Technology of China, by NSFC grants NSFC-11773029,
-11833008, -11991054, -11922304, -11873048, -11690024, -11703077,
by the CAS Key Research Program through KJZD-EW-M06, by the Key Research
Program of Frontier Sciences, CAS, grant QYZDJ-SSW-SLH007.}

\appendix

\section{CCF analysis in each velocity bin}

\begin{figure}
  \centering
  \includegraphics[width=0.45\textwidth]{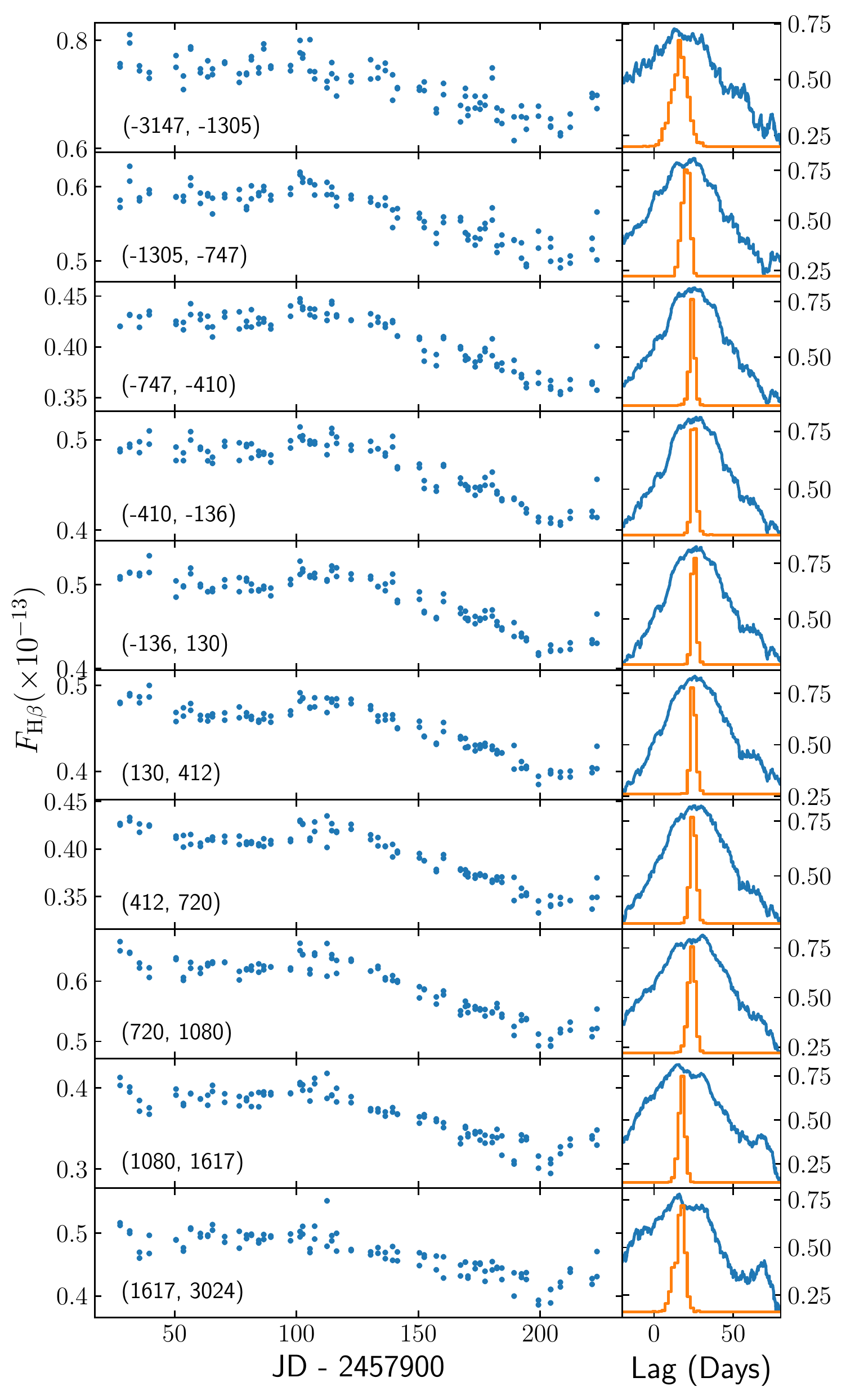}
  \includegraphics[width=0.45\textwidth]{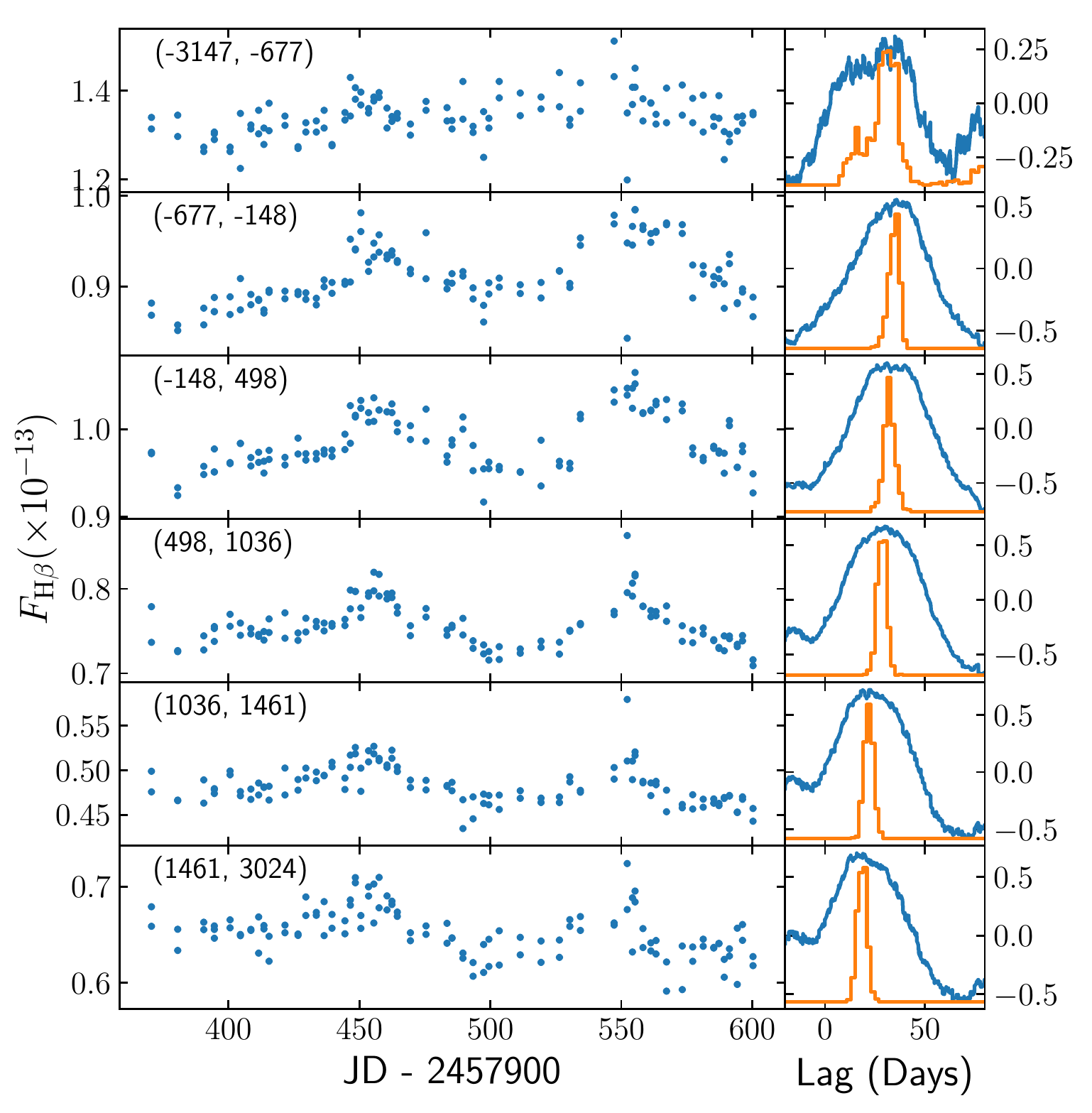}
  \caption{\hb\ light curves and CCF analysis in each velocity bin with equal
  flux in the rms spectrum, from 2017 (left) and 2018 (right) observations.
  The blue curves are CCF functions and yellow lines are Monte Carlo
  simulations as described in the main text. The velocity boundaries are given
  by numbers in the brackets.}
  \label{fig-velocity-bin_rms}
\end{figure}

Figure \ref{fig-velocity-bin_rms} shows \hb\ light curves, CCFs, and CCCDs in
each velocity bin with equal flux in the rms spectrum, used in
Figure \ref{velocity-resolved}.

\section{Velocity-resolved delays with equal fluxes in the mean spectra}

\begin{figure}
  \centering
  \includegraphics[width=0.4\textwidth]{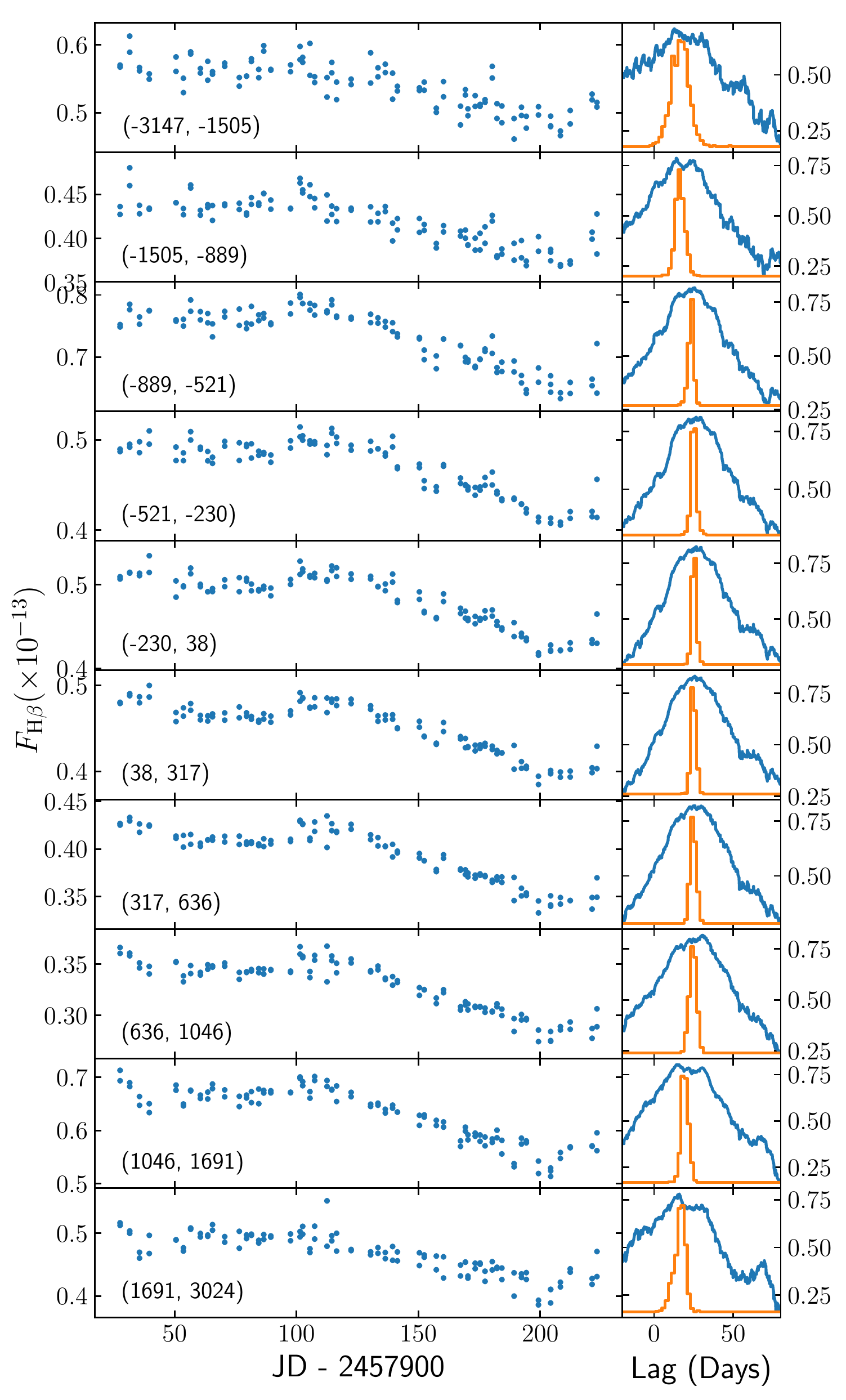}
  \includegraphics[width=0.4\textwidth]{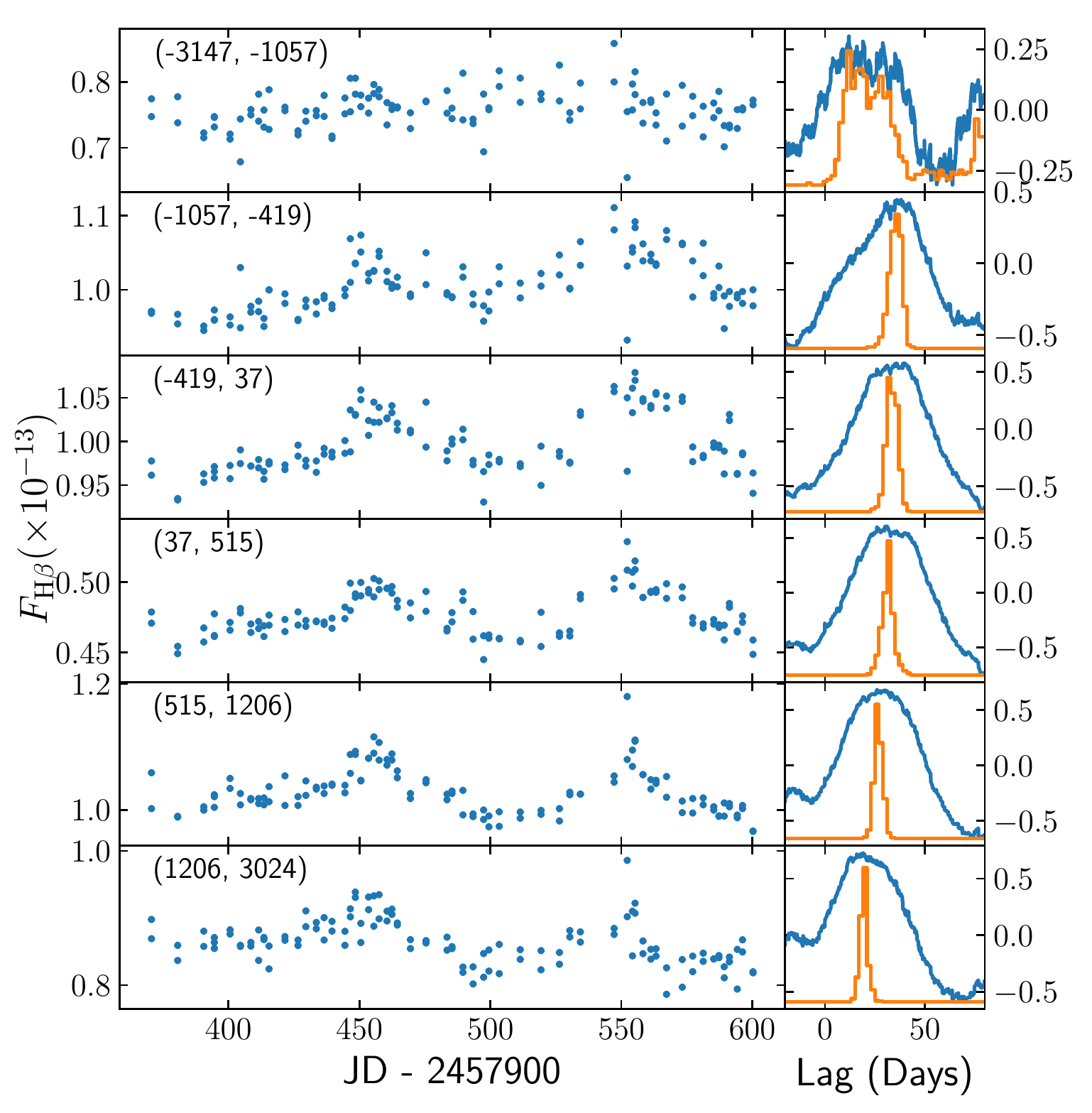}
  \caption{\hb\ light curves and CCF analysis in each velocity bin with equal
  flux in the mean spectrum, from 2017 (left) and 2018 (right) observations.
  The blue curves are CCF functions and yellow lines are Monte Carlo
  simulations. The velocity boundaries are given in the brackets.}
  \label{fig-velocity-bin_mean}
\end{figure}

\begin{figure}
  \centering
  \includegraphics[width=0.4\textwidth]{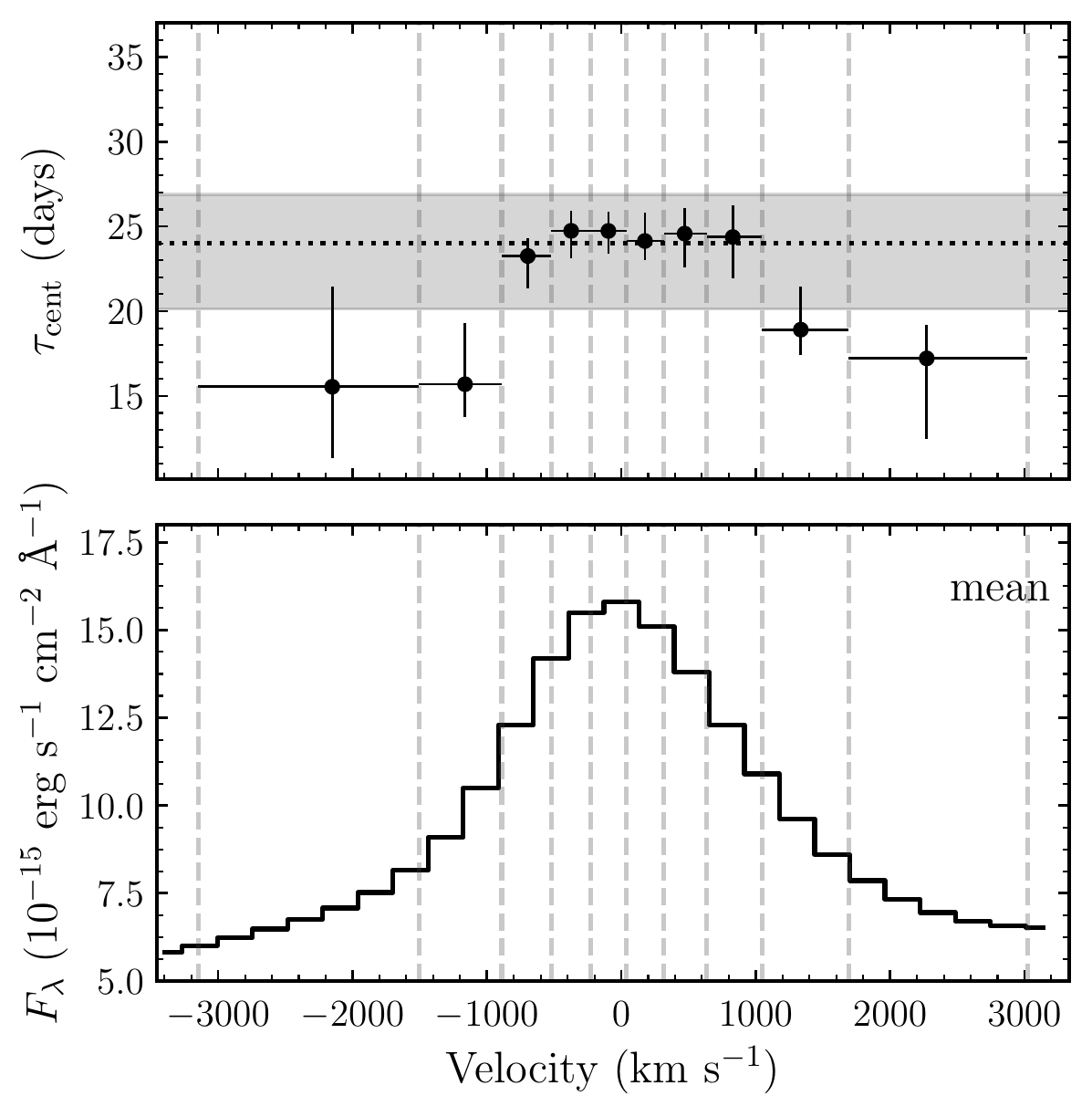}
  \includegraphics[width=0.4\textwidth]{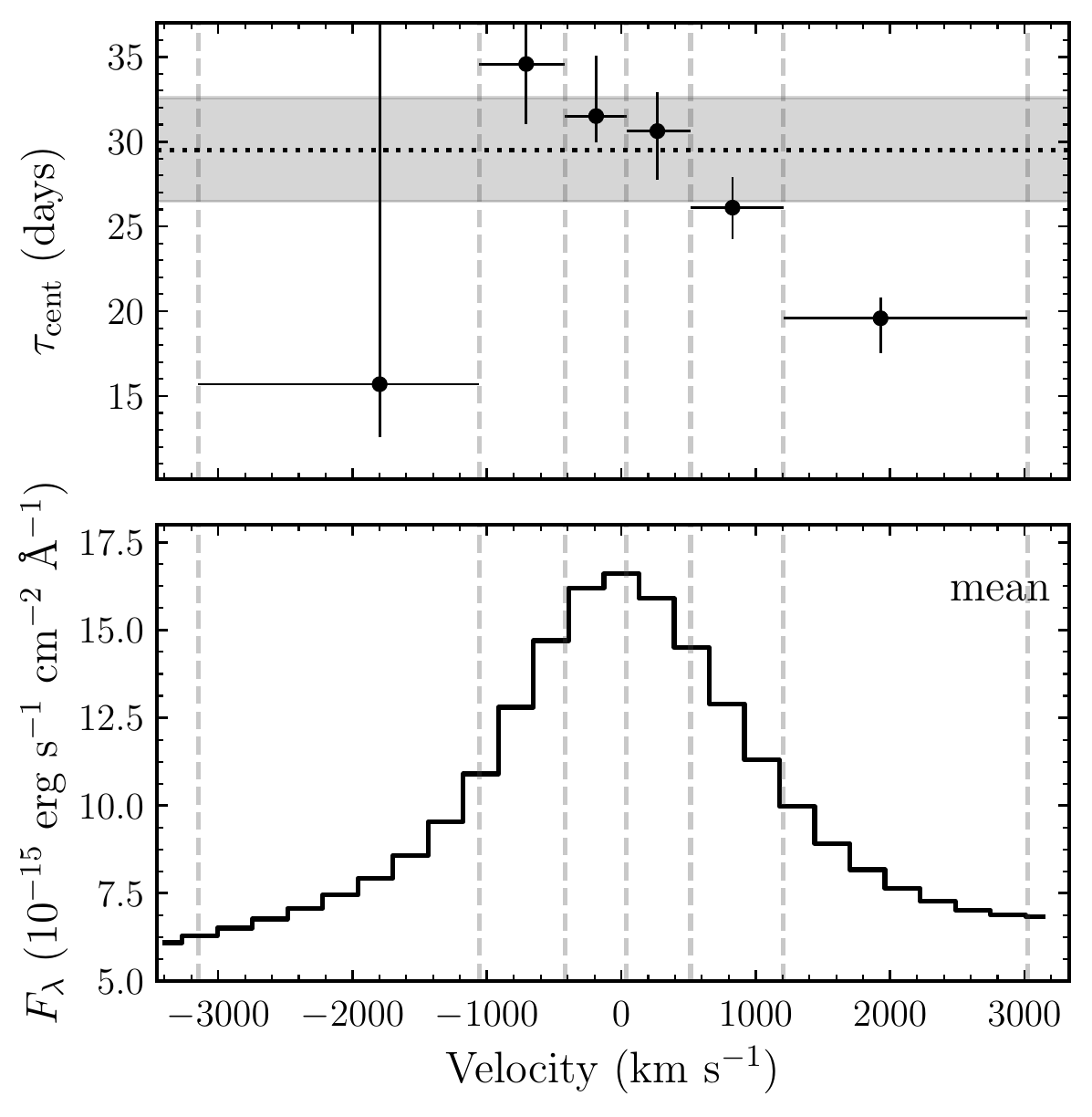}
  \caption{Velocity-resolved delays in each velocity bin with equal flux in
  the mean spectrum, for 2017 (left) and 2018 (right) observations.}
  \label{fig-velocity-resolved_mean}
\end{figure}

Considering that the rms spectrum could be contaminated by continuum, we
calculated the \hb\ velocity-resolved delays in each velocity bin with equal
flux in the mean spectrum, shown in Figure \ref{fig-velocity-resolved_mean}.
Light curves in each velocity bin are given in Figure
\ref{fig-velocity-bin_mean}. Comparing Figure \ref{velocity-resolved} with
\ref{fig-velocity-resolved_mean}, we find that the two kinds of
velocity-resolved delays are quite similar. Therefore the influence of
continuum contamination is not important. 

\end{document}